\documentclass[twocolumn]{aastex631}

\usepackage{amsmath}
\usepackage{amssymb}
\usepackage{graphicx}

\newcommand{\fNL}{f_\mathrm{NL}}
\newcommand{\fNLloc}{f_\mathrm{NL}^\mathrm{local}}
\newcommand{\fNLeq}{f_\mathrm{NL}^\mathrm{equil}}
\newcommand{\fNLort}{f_\mathrm{NL}^\mathrm{ortho}}

\newcommand{\kmin}{k_\mathrm{min}}
\newcommand{\kmax}{k_\mathrm{max}}
\newcommand{\Ngrid}{N_\mathrm{grid}}
\def\bk{\mathbf{k}}

\def\bx{\mathbf{x}}

\def\d{\mathrm{d}}

\newcommand{\hMpc}{h\,\mathrm{Mpc}^{-1}}

\def\btheta{\boldsymbol{\theta}}
\def\t{\mathbf{t}}
\def\cov{\mathbf{C}}
\def\mean{\boldsymbol{\mu}}
\def\data{\mathbf{d}}
\def\transpose{\mathrm{T}}
\def\fisher{\mathbf{F}}

\newcommand{\Quijote}{\textsc{Quijote}}
\newcommand{\QuijotePNG}{\textsc{Quijote-png}}

\newcommand{\cnrs}{Sorbonne Universit\'{e}, CNRS, UMR 7095, Institut d'Astrophysique de Paris, 98 bis bd Arago, 75014 Paris, France}
\newcommand{\cca}{Center for Computational Astrophysics, Flatiron Institute, 162 5th Avenue, New York, NY 10010, USA}
\newcommand{\bologna}{Dipartimento di Fisica e Astronomia, Alma Mater Studiorum - University of Bologna, Via Piero Gobetti 93/2, 40129 Bologna BO, Italy}
\newcommand{\inaf}{INAF - Osservatorio Astronomico di Bologna, Via Piero Gobetti 93/3, 40129 Bologna BO, Italy}
\newcommand{\infn}{INFN - Istituto Nazionale di Fisica Nucleare, Sezione di Bologna, Viale Berti Pichat 6/2, 40127 Bologna BO, Italy
}
\newcommand{\infnPad}{INFN, Sezione di Padova, via Marzolo 8, I-35131, Padova, Italy}
\newcommand{\mpa}{Max-Planck-Institut f\"ur Astrophysik, Karl-Schwarzschild-Straße 1, 85748 Garching, Germany}
\newcommand{\ICC}{ICC, University of Barcelona, IEEC-UB, Martí i Franquès, 1, E-08028 Barcelona, Spain}
\newcommand{\ICREA}{ICREA, Pg. Lluís Companys 23, Barcelona, E-08010, Spain}
\newcommand{\Galilei}{Dipartimento di Fisica e Astronomia “G. Galilei”,Università degli Studi di Padova, via Marzolo 8, I-35131, Padova, Italy}
\newcommand{\uwc}{Department of Physics and Astronomy, University of the Western Cape, Cape Town 7535, South Africa}
\newcommand{\princeton}{Department of Astrophysical Sciences, Princeton University, 4 Ivy Lane, Princeton, NJ 08544 USA}

\begin{document}

\title{Quijote-PNG: Quasi-maximum likelihood estimation of Primordial Non-Gaussianity in the non-linear dark matter density field}

\author{Gabriel Jung}
\affiliation{\Galilei}
\affiliation{\infnPad}
\author{Dionysios Karagiannis}
\affiliation{\uwc}
\author{Michele Liguori}
\affiliation{\Galilei}
\affiliation{\infnPad}
\author{Marco Baldi}
\affiliation{\bologna}
\affiliation{\inaf}
\affiliation{\infn}
\author{William R Coulton}
\affiliation{\cca}
\author{Drew Jamieson}
\affiliation{\mpa}
\author{Licia Verde}
\affiliation{\ICREA} 
\affiliation{\ICC}
\author{ Francisco Villaescusa-Navarro}
\affiliation{\cca}
\affiliation{\princeton}
\author{Benjamin D. Wandelt}
\affiliation{\cnrs}
\affiliation{\cca}

\begin{abstract}
   Future Large Scale Structure surveys are expected to improve over current bounds on primordial non-Gaussianity (PNG), with a significant impact on our understanding of early Universe physics. The level of such improvements will however strongly depend on the extent to which late time non-linearities erase the PNG signal on small scales. In this work, we show how much primordial information remains in the bispectrum of the non-linear dark matter density field by implementing a new, simulation-based, methodology for joint estimation of primordial non-Gaussian amplitudes ($\fNL$) and standard $\Lambda$CDM parameters. The estimator is based on optimally compressed statistics, which, for a given input density field, combine power spectrum and modal bispectrum measurements, and numerically evaluate their covariance and their response to changes in cosmological parameters. 
   In this first analysis, we focus on the matter density field, and we train and validate the estimator using a large suite of N-body simulations (\QuijotePNG), including different types of PNG (local, equilateral, orthogonal). We explicitly test the estimator's unbiasedness, optimality and stability with respect to changes in the total number of input realizations. While the dark matter power spectrum itself contains negligible PNG information, as expected, including it as an ancillary statistic increases the PNG information content extracted from the bispectrum by a factor of order $2$. As a result, we prove the capability of our approach to optimally extract PNG information on non-linear scales beyond the perturbative regime, up to $k_{\rm max} = 0.5~h\,{\rm Mpc}^{-1}$, obtaining marginalized $1$-$\sigma$ bounds of $\Delta f_{\rm NL}^{\rm local} \sim 16$, $\Delta f_{\rm NL}^{\rm equil} \sim 77$ and $\Delta f_{\rm NL}^{\rm ortho} \sim 40$ on a cubic volume of $1~(\mathrm{Gpc}/h)^3$ at $z=1$. At the same time, we discuss the significant information on cosmological parameters contained on these scales.
\end{abstract}

\keywords{primordial non-Gaussianity, large scale structure, optimal estimator, bispectrum}

\section{Introduction}\label{sec:intro}

The best cosmological probe of primordial non-Gaussianity (PNG) has been, up to now, the Cosmic Microwave Background (CMB) \citep{Planck:2019kim}. The available information in the angular bispectrum (i.e., the three-point function of harmonic multipoles) of CMB primary anisotropies has been, however, nearly completely extracted. If we want to obtain significant improvements over current primordial NG (PNG) constraints for most models, we thus have to look at different observables. To this purpose, galaxy clustering carries a significant potential \citep{Alvarez:2014vva, Karagiannis:2018jdt, Meerburg:2019qqi}, since the 3D galaxy density field contains many more bispectrum modes than the 2D CMB fluctuation fields. Most of these modes are however in the gravitational non-linear regime. The important challenge to face is therefore to separate, up to the smallest possible scales, the PNG signal from the late-time NG component, induced by evolution of cosmic structures. 

To tackle this issue, significant efforts have been devoted to obtain theoretical predictions of the galaxy bispectrum, via a suitable perturbative treatment valid on the largest scales \citep[for the current state of the art, see][and references therein]{Ivanov:2021kcd}. This approach has even led recently to the first constraints on PNG using both the galaxy power spectrum and bispectrum, by analyzing data from the BOSS survey \citep{2013AJ....145...10D, Cabass:2022wjy, Cabass:2022ymb, DAmico:2022gki}.\footnote{Previous analyses based on the BOSS galaxy bispectrum, in different cosmological contexts, can be found for example in \citet{Gil-Marin:2014sta, Gil-Marin:2014baa, Gil-Marin:2016wya, 2017MNRAS.469.1738S, 2017MNRAS.468.1070S}.} Recent works based on field-level inference \citep{Baumann:2021ykm, Andrews:2022nvv}, neural networks \citep{Giri:2022nzt}, or reconstruction methods \citep{Shirasaki:2020vkk}, have however shown that much more information should be present in the data. To extract this extra information, alternative promising observables as the density probability density function \citep{Mao:2014caa, Uhlemann:2017tex, Friedrich:2019byw}, persistent homology \citep{Biagetti:2020skr, Biagetti:2022qjl}, or higher order correlation functions in Fourier space \citep{Gualdi:2020eag, Gualdi:2021yvq}, have been considered. It has also been confirmed that probing non-linear scales, using the power spectrum and the bispectrum, improves significantly the constraints on cosmological parameters \citep{Hahn:2019zob, Hahn2}. Checking if this is also the case for PNG is then an important question. It is however unclear if analytical approaches can go significantly beyond the mildly non-linear regime.

In this work, we thus consider instead the alternative, simulation-based route using a large number of realizations of the matter density field. This paper, and its companion paper \citet{Coulton:2022}, are the first of a series of studies, in which we plan to take a step-by-step approach and address the problem in increasing level of complexity. We start here by discussing the general implementation of our statistical estimation and data compression algorithms based on the modal bispectrum estimator, while, in \citet{Coulton:2022} we describe in more detail our input simulations and use them to perform an independent Fisher matrix analysis using a binned bispectrum estimator. Moreover, for additional complementarity, the two analyses are performed at different redshifts. In \citet{Coulton:2022}, the main focus is on $z=0$, where the non-linear effects are maximal. Here, we study the redshift $z=1$, more suited for comparison with upcoming surveys such as Euclid \citep{2011arXiv1110.3193L, Amendola:2016saw}. At this initial stage, we focus our analysis just on the dark matter field, with a twofold aim: to setup the general pipeline that we will also use in following analyses and to address the first crucial question of how much PNG information can be in principle extracted from the matter field, by pushing the analysis up to small scales, which would be hard to model analytically. In a follow up paper, we will extend this analysis to halos, whereas in later studies we will finally consider the galaxy density field and account for additional important effects, such as redshift-space distortions and incomplete sky coverage. 

We extract the power spectrum and bispectrum of a large number of realizations of the matter density field, allowing us to describe precisely the contribution of PNG down to nonlinear scales ($\kmax=0.5~\hMpc$), as well as measuring the corresponding covariance matrix including all non-Gaussian and off-diagonal terms \citep[see][for a discussion of their relative importance]{Biagetti:2021tua}. We then combine these measurements into optimally compressed statistics, a procedure which was shown to be extremely efficient to deal with the galaxy bispectrum in \citet{Gualdi:2017iey, Gualdi:2018pyw}. Finally, following the general scheme developed in \citet{Alsing:2017var}, we derive an estimator to jointly measure $\Lambda$CDM and PNG parameters.

The plan of the paper is as follows. In section \ref{sec:primordialNG} we describe the PNG models that we consider in this work and define the parameters in our analysis; in section \ref{sec:estimators} we describe our estimators of the power spectrum and bispectrum and the optimal data compression procedure applied to build a quasi-maximum likelihood estimator of PNG and cosmological parameters; in section \ref{sec:analyses} we discuss the application of these estimators to our input mock datasets and show the main results of our analysis; in section \ref{sec:conclusions} we summarize our conclusions and discuss future prospects. 
\section{Primordial bispectrum shapes}
\label{sec:primordialNG}

For most inflationary models, the main PNG signature is a non-vanishing bispectrum in the primordial curvature perturbation field, which is defined as:
\begin{equation}
\begin{split}
\langle \Phi(\mathbf{k_1}) \Phi(\mathbf{k_2}) \Phi(\mathbf{k_3}) \rangle = (2 \pi)^3 \fNL & F(k_1, k_2, k_3) \\ &\times \delta^{(3)}(\mathbf{k_1} + \mathbf{k_2} + \mathbf{k_3}) ,
\end{split}
\end{equation}
where $\Phi$ is related to the comoving perturbation variable on super-horizon scales by $\Phi \equiv \frac{3}{5} \zeta$ and the Dirac delta function enforces translation invariance. The quantity $F(k_1,k_2,k_3)$ defines instead the functional dependence of the bispectrum on different Fourier space triplets -- the so-called bispectrum shape -- and it depends only on the length of the wavevectors in virtue of rotation invariance. Finally, the dimensionless PNG amplitude parameter $\fNL$ measures the strength of the PNG signal, for a given shape. A main goal in our analysis is to build an efficient $\fNL$ estimator for the three main bispectrum shapes, namely the local, equilateral and orthogonal\footnote{In the companion paper \citep{Coulton:2022}, two different orthogonal shapes are discussed; the standard CMB one, which we use here, and a corrected template more accurate for LSS. While this may change the numerical results presented in this work, as they have different behaviours in the squeezed limit, it has no impact on the general conclusions.} shapes, defined by the following templates \citep[see][and references therein]{Planck:2019kim}:
\begin{eqnarray}
\label{eqn:localbis}
F^{\rm local}(k_1,k_2,k_3) & = & 2 A^2 \left[ \frac{1}{k_1^{4-n_s}} \frac{1}{k_2^{4-n_s}} + \rm{cycl.} \right] ,
\end{eqnarray}
\begin{eqnarray}
\label{eqn:equilateralbis}
\nonumber
& &	F^{\rm equil}(k_1,k_2,k_3)= 6A^2 \\
\nonumber
& \times& \left\{
\left[ -\frac1{k^{4-n_{\rm s}}_1k^{4-n_{\rm s}}_2} - {\rm cycl.} \right]
-\frac2{(k_1k_2k_3)^{2(4-n_{\rm s})/3}}
\right. \\
& &
\left.+\left[\frac1{k^{(4-n_{\rm s})/3}_1k^{2(4-n_{\rm s})/3}_2k^{4-n_{\rm s}}_3}
+\mbox{(5 perms.)}\right]\right\}\, , 
\end{eqnarray}
\begin{eqnarray}
\label{eqn:orthogonalbis}
\nonumber
& &	F^{\rm ortho}(k_1,k_2,k_3)= 6A^2 \\
\nonumber
& \times& \left\{
\left[ -\frac3{k^{4-n_{\rm s}}_1k^{4-n_{\rm s}}_2} - {\rm cycl.} \right]
-\frac8{(k_1k_2k_3)^{2(4-n_{\rm s})/3}}
\right. \\
& &
\left.+\left[\frac3{k^{(4-n_{\rm s})/3}_1k^{2(4-n_{\rm s})/3}_2k^{4-n_{\rm s}}_3}
+\mbox{(5 perms.)}\right]\right\}\,. 
\end{eqnarray}

In the present study we directly work with the matter density field; in such case, for small PNG, nearly all the information on $\fNLloc$, $\fNLeq$, $\fNLort$ is contained in the field bispectrum. When dealing instead with observed dark matter tracers, it is well known that an important additional signature of PNG arises in the form of a scale dependent bias term, sourced by the squeezed configurations of the primordial bispectrum \citep{Dalal2008,Slosar2008,Matarrese2008,Afshordi2008,Verde2009,Desjacques2010} (hence, especially important for PNG of the local type). In such case, the large scale power spectrum carries significant extra information on the PNG amplitude. Since at this stage we perform joint estimation of $f_{\rm NL}$ and standard $\Lambda$CDM cosmological parameters, the power spectrum is already included in our analysis. Therefore, in follow-up studies that will consider dark matter halos and galaxies, no essential modification of our current data analysis pipeline will be needed, except in the input fields to analyze.   
\section{Estimators}
\label{sec:estimators}

\subsection{Summary statistics}
\label{sec:summary}

As discussed, the aim of our work is to efficiently combine power spectrum and bispectrum measurements of the matter field in the input simulations, to maximize the PNG information that can be extracted, up to the smallest possible scales. This will be achieved by resorting to a suitable compression scheme, taking the power spectrum and bispectrum as the starting summary statistics and requiring numerical evaluation of their covariances, to build the standard quasi-maximum likelihood estimator described in \citet{Alsing:2017var}. This data compression step is also a key ingredient to perform density-estimation likelihood-free inference \citep[see e.g.][]{Alsing:2019xrx}, an alternative approach we aim to pursue in a future work. Forecasts based on data compression also have the advantage of being conservative, whereas a standard Fisher matrix analysis using the starting set of modes could in principle underestimate errors; this would happen in case of a lack of statistical convergence in the Monte Carlo averages, due to a low number of input simulations. In the current analysis, we however explicitly show that we can achieve optimality, thanks to fast and efficient estimators to measure the power spectra and bispectra, which are applied to a large enough set of input realizations.

\subsubsection{Power spectrum}

To measure the power spectrum, we use the standard estimator \citep[e.g.][]{Feldman:1993ky}
\begin{equation}
    \label{eq:P(k)}
    \hat{P}(k_i) = \frac{1}{ V N_i} \sum\limits_{\bk \in \Delta_i} \delta(\bk)\delta^*(\bk),
\end{equation}
where $\delta(\bk)$ is the density field in Fourier space defined on a grid, the $k$-range has been divided into bins $\Delta_i$ of width the fundamental mode, $V$ is the survey volume and $N_i$ is the number of vectors $\bk$ in each bin.

\subsubsection{Modal bispectrum}

 A well suited choice to measure the bispectrum statistics is the so called modal bispectrum estimator, originally developed for CMB NG analysis in \citet{Fergusson:2009nv, Fergusson:2010dm} and then extended to LSS in \citet{Fergusson:2010ia, Regan:2011zq, Schmittfull:2012hq} \citep[see also][]{Lazanu:2015rta, Lazanu:2015bqo,Hung:2019ygc, Hung:2019nma, Byun:2020rgl, Byun:2022rvn}.

The main idea of modal estimation is to expand a general weighted bispectrum shape in a factorizable basis:
\begin{equation}
\label{eq:modal-bispectrum}
w(k_1, k_2, k_3) B(k_1, k_2, k_3)
\simeq \sum_{m,n=0}^{N-1} \gamma^{-1}_{mn}\beta_m Q_n(k_1, k_2, k_3),  
\end{equation}
where $Q_n \equiv q_{\{p}(k_1) q_r(k_2) q_{s \}}(k_3)$ corresponds to the symmetrized product of one-dimensional basis functions $q_p(k)$ and a partial ordering $n \leftrightarrow (p,q,r)$ has been introduced \citep[see e.g.][]{Fergusson:2010dm}, $w$ is a weight function and $\beta$ are the expansion coefficients. If the input bispectrum is a smooth function of $k_1,k_2,k_3$, it can be well approximated by truncating the sum over a number $N$ of modes that is much smaller than the total number of Fourier triangles. $\gamma_{mn}\equiv\langle Q_m | Q_n \rangle$ is the inner product between mode functions, where the following scalar product (see appendix~\ref{app:gamma} for details on the derivation) 
\begin{align}\label{eq:inner}
    \left\langle f(k_1,k_2,k_3) | g(k_1,k_2,k_3) \right \rangle & \equiv \frac{1}{8\pi^4}\int_{\cal{V}_T} \d k_1 \d k_2 \d k_3 \\ & \times f(k_1, k_2,k_3) g(k_1,k_2,k_3),
\end{align}
is computed over the bispectrum \emph{tetrapyd} domain $\cal{V}_T$, i.e., the set of all the Fourier triplets that form a closed triangle and such that each wavenumber has a length in the chosen range [$\kmin$, $\kmax$].  

Each coefficient $\beta_n$ can be estimated by fitting the corresponding mode $Q_n$ to the data bispectrum. Making the customary choice of a separable weight function using the total power spectrum $P(k)$:
\begin{equation}
    w(k_1, k_2, k_3) \equiv \frac{\sqrt{k_1 k_2 k_3}}{\sqrt{P(k_1)P(k_2)P(k_3)}},
\end{equation}
one can show e.g., \citep[e.g.][]{Fergusson:2010ia} that the modal estimator takes the form:
\begin{equation}
    \label{eq:beta}
    \hat{\beta}_n = \frac{1}{V}\int \d^3 x\, M_p(\bx)M_q(\bx)M_r(\bx),
\end{equation}
where
\begin{equation}
    \label{eq:M}
    M_p(\bx) \equiv \int \frac{\d^3 k}{(2\pi)^3} \frac{q_p(k)\delta(\bk)}{\sqrt{k P(k)}}e^{i\bk.\bx}.
\end{equation}
This inverse Fourier transform can be computed in an efficient way using fast Fourier transformation (FFT) routines (e.g. FFTW\footnote{\url{http://www.fftw.org}}), making $\hat{\beta}_n$ extremely fast to extract from data. In this paper, we work directly with the observed modal coefficients $\beta_n$ to estimate primordial non-Gaussianity amplitudes $\fNL$ and other cosmological parameters. We therefore do not need to evaluate the inner product matrix $\gamma$, a rather technical step of the numerical pipeline. The only part in this work, for which the calculation of $\gamma$ is needed is when we reconstruct the bispectrum from the modal basis, presented in figures~\ref{fig:bisp-png} and \ref{fig:bisp-cosmo}, obtained by substituting the measured $\beta_n$ into eq.~\eqref{eq:modal-bispectrum}. We describe our method  to compute $\gamma$ in detail in appendix~\ref{app:gamma}.

The basis of one-dimensional functions $q_p(k)$ we use here is mainly composed of well-behaved polynomials normalised within the \emph{tetrapyd} that fits inside a unit cube \citep[see][for details]{Fergusson:2010dm}. This means that in the above expressions $q_p(k)\rightarrow q_p((k-\kmin)/(\kmax-\kmin))$. We also include custom modes based on the separable matter bispectrum shapes, i.e.~the gravity induced tree-level bispectrum and the local PNG template eq.~\eqref{eqn:localbis} to further improve the bispectrum decomposition in eq.~\eqref{eq:modal-bispectrum}. This was introduced in \citet{Hung:2019ygc} and further developed in \citet{Byun:2020rgl}, see appendix~\ref{app:basis} for details. A general useful feature of the modal estimator is that, for a proper choice of basis, it allows for a preliminary data compression step, by reconstructing the signal in a small number of modes, compared to the starting amount of triplets in Fourier space. This speeds up the numerical evaluation of covariances, which is in turn a key ingredient for the final compression step described in the next section.

\subsection{Data compression and quasi maximum-likelihood estimator}
\label{sec:compression}

To build an optimal estimator of primordial non-Gaussianity amplitude parameter $\fNL$'s, we follow the method developed in \citet{Alsing:2017var}, based on a two-step data compression scheme. The first step consists on extracting a set of summary statistics from a full dataset, which in our case are the power spectrum $P(k)$ and bispectrum modes $\beta_n$, described in previous paragraphs. The second step is an additional compression, down to only $n$ numbers where $n$ is the number of parameters we aim to infer from data. The optimal compressed quantities, in the sense that no information about the wanted parameters is lost, are the components of the score function (the parameter gradient of the log-likelihood), which can be computed explicitly by assuming an approximate form for the likelihood of the summary statistics. When working with the power spectrum and bispectrum, a Gaussian likelihood is a reasonable approximate choice, by virtue of the central limit theorem.\footnote{The validity of this assumption is tested a posteriori with the different analyses presented in section~\ref{sec:analyses}, where we verify that the estimator developed yields unbiased and quasi-optimal results.} The compressed statistics , called $\t$, then take the following form:
\begin{equation}
    \t = \nabla_{\btheta}\mean_*^\transpose\cov^{-1}_*(\data - \mean_*),
    \label{eq:compression}    
\end{equation}
where $\data$ corresponds to the summary statistics (a vector of modal coefficients $\beta_n$ and possibly power spectrum bins $P(k)$), $\mean$ and $\cov$ are respectively the mean and the covariance of $\data$, $\btheta$ denotes the parameters of interest and the subscript $*$ indicates that the quantity is evaluated at the chosen fiducial cosmology. Note that in eq.~\eqref{eq:compression}, we made the further assumption that the covariance $\cov$ does not depend on the parameters $\theta$ and in that case the compressed statistics $\t$ is equivalent to the {\sc moped} data compression of \citet{Heavens:1999am}. By construction, the covariance of the compressed statistic $\t$ at the fiducial point $\btheta_*$ is equal to the Fisher matrix $\fisher_*$, where 
\begin{equation}
    \fisher = \nabla_{\btheta}\mean^\transpose \cov^{-1} \nabla^\transpose_{\btheta}\mean.
    \label{eq:fisher}
\end{equation}

One can also write the following quasi maximum-likelihood estimator for the parameters $\btheta$:
\begin{equation}
    \hat\btheta = \btheta_* + \fisher^{-1}_*\t,
    \label{eq:ml-estimator}
\end{equation}
which has an expected covariance given by the Fisher matrix evaluated at the fiducial point $\fisher_*$.

In this work, we will compute both the derivatives and the covariance matrix from N-body simulations, which makes it possible to include nonlinear scales in the analysis and estimate their constraining power on primordial non-Gaussianity. 
\section{Analysis}
\label{sec:analyses}

\subsection{Simulations}
\label{sec:simulations}

The analysis presented in this work is mainly based on the publicly available \Quijote\ suite of N-body simulations,\footnote{\url{https://quijote-simulations.readthedocs.io}} see \citet{Villaescusa-Navarro:2019bje} for details. These simulations are cubic boxes of length $1~h^{-1}$Gpc, containing $512^3$ particles, generated with the \textsc{Gadget-III} code \citep{Springel:2005mi}, with input transfer functions computed by CAMB \citep{Lewis:1999bs}. Initial conditions are generated at $z_i=127$ using the code \textsc{2LPTIC} \citep{Crocce:2006ve}. We use a set of $8,000$ simulations at fiducial cosmology to evaluate the covariance matrix, together with smaller sets of $500$ realizations, in which parameters are one by one slightly displaced from their fiducial values, to numerically compute the derivatives in eq.\ \ref{eq:compression}. Fiducial parameter values and steps used for numerical differentiation are reported in table~\ref{tab:quijote}. In appendix~\ref{app:ratio-params}, we illustrate the effects of small parameter variations on the power spectrum and bispectrum. Throughout the analysis, we work at redshift $z=1$. 

\begin{deluxetable*}{c|cccccccc}[]
\tablecaption{The cosmological parameters and PNG amplitudes of the \Quijote~and \QuijotePNG~simulations.  \label{tab:quijote}}
\tablehead{& $\sigma_8$ & $\Omega_m$ & $\Omega_b$ & $n_s$ & $h$ & $\fNLloc$ & $\fNLeq$ & $\fNLort$}
\startdata
Fiducial & 0.834 & 0.3175 & 0.049 & 0.9624 & 0.6711 & 0 & 0 & 0 \\
    Steps & $\pm 0.015$ & $\pm 0.01$ & $\pm 0.002$ & $\pm 0.02$ & $\pm 0.02$ & $\pm 100$ & $\pm 100$ & $\pm 100$ \\
\enddata
\end{deluxetable*}

To study the effects of PNG, we use the \QuijotePNG\ set presented in detail in the companion paper \citep{Coulton:2022}. Non-Gaussian initial conditions are generated using the method developed in \citet{Scoccimarro:2011pz}, and evolved following the exact same procedure as the standard \Quijote\ N-body simulations described above. For each of the three standard primordial shapes -- local (eq.\ \ref{eqn:localbis}), equilateral (eq.\ \ref{eqn:equilateralbis}) and orthogonal (eq.\ \ref{eqn:orthogonalbis}) -- we analyze two sets of $500$ realizations, with either $\fNL=+100$ or $\fNL=-100$. 

For further testing, we also consider three additional sets of $10$ simulations, with $\fNLloc=-40$, $0$ and $40$, all generated from the same Gaussian seeds, independently of the \Quijote\ fiducial realizations, but with the same numerical specifications and input cosmological parameters. For this suite of simulations, initial conditions have been generated with a modified version of the \texttt{PNGRun} code \citep{Wagner:2010me} that imprints local NG contributions on top of the primordial Gaussian potential field, starting from a prescribed shape of the primordial matter bispectrum \citep[see][for a detailed description of the \texttt{PNGRun} algorithm]{Wagner:2010me}. The simulations have been evolved with \textsc{Gadget-III} by including the contribution of relativistic species in the background cosmic expansion, which results in a different normalisation at the starting redshift of the runs, compared to \Quijote\, to match the same perturbations amplitude at $z=0$. 

\subsection{Results}
\label{sec:measurements}

The first step of our analysis consists in the extraction of power spectrum and bispectrum modes from all available simulations, using the methodology outlined in section~\ref{sec:summary}.

Modal bispectrum estimation requires us to fix $\kmax$ beforehand. In this work, we consider three different values: $\kmax=0.07~\hMpc$ (linear), $0.2~\hMpc$ (mildly non-linear) and $0.5~\hMpc$ (non-linear), in order to study the amount of information as a function of scales. We begin by depositing particle positions into a regular grid using the public Pylians3\footnote{\url{https://github.com/franciscovillaescusa/Pylians3}} code, with a fourth-order interpolation scheme and a resolution $\Ngrid=360$ in the non-linear case, as in \citet{Hahn:2019zob}, and the faster cloud-in-cell interpolation at the lower resolutions $\Ngrid=128$ and $\Ngrid=64$\footnote{It has been argued in \citet{Sefusatti:2015aex} that the FFT-based standard bispectrum estimator can probe scales up to to $\kmax=2k_\mathrm{Ny}/3$, where the Nyquist frequency is $k_\mathrm{Ny}=k_f \Ngrid/2$ and $k_f=2\pi/L$ is the fundamental frequency of the box. This is due to the factor $e^{i(\bk_1+\bk_2+\bk_3)\cdot x}$, present in the standard estimator, which is invariant under a translation $k_i\rightarrow k_i+k_f \Ngrid/3$. This indicates that any estimator that has this exponent factor will have the above momentum cutoff scale, rather than $\kmax=k_\mathrm{Ny}$. In \citet{Byun:2020rgl} they test this and find indications that the modal estimator falls under the same category, since it takes a form similar to the standard bispectrum estimator (see eq.\ \ref{eq:inner_3d}). Taking this into account, gives the largest values for the wavenumbers k to be $\kmax \simeq 0.75$, $0.26$ and $0.13~\hMpc$ for $\Ngrid=360$, $128$ and $64$ respectively, well above our chosen $\kmax$ values.} for the $\kmax=0.2$ and $0.07~\hMpc$ cases respectively.

\begin{figure}
    \includegraphics[width=0.99\linewidth]{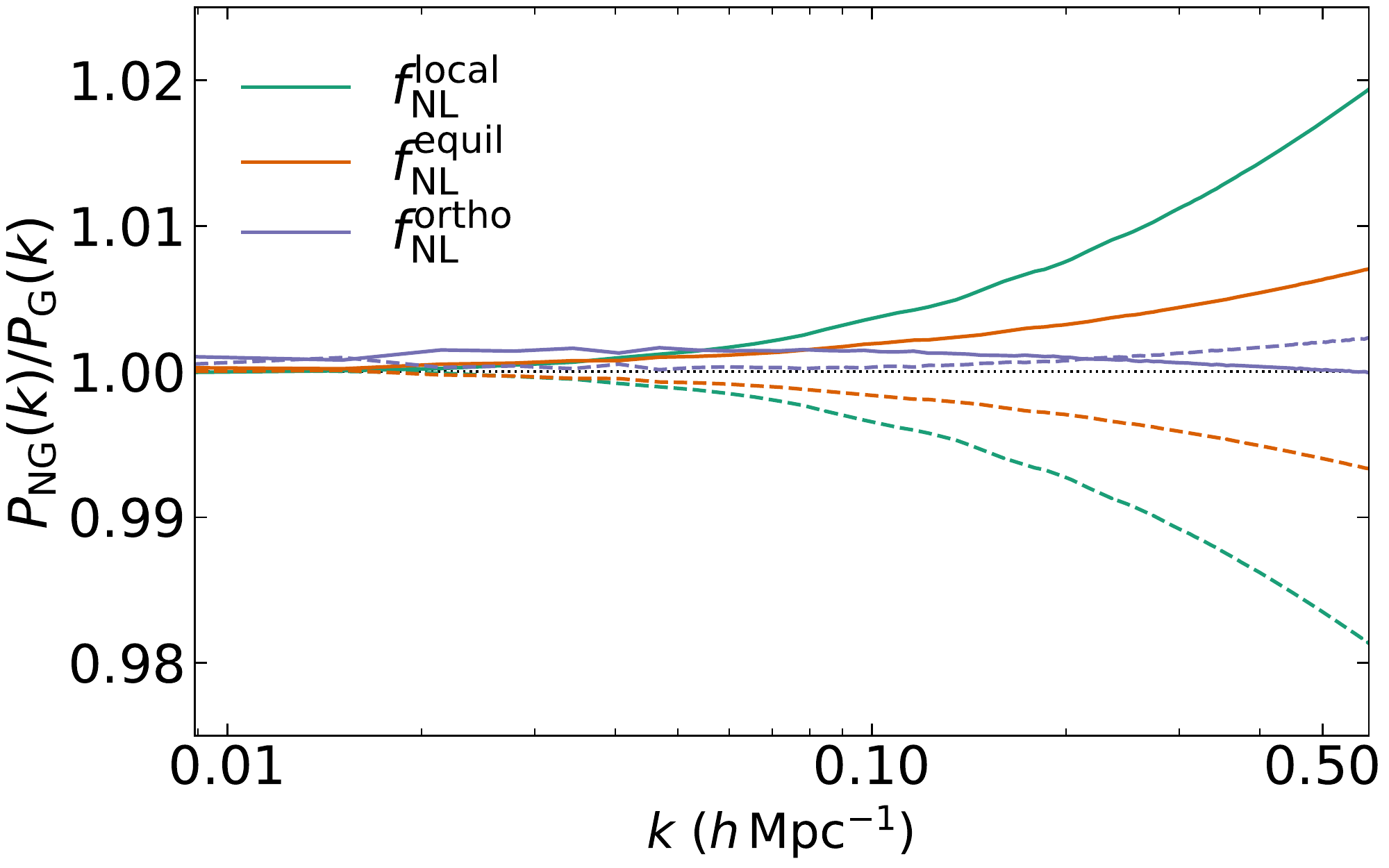}
    \caption{The impact of PNG on the matter power spectrum at $z=1$ (ratio NG/G, averaged from $500$ N-body simulations). We show the local (green), equilateral (orange), and orthogonal (purple) shapes. Solid and dashed lines correspond respectively to positive and negative $\fNL$ values ($\pm100$).}
    \label{fig:pk-png}
\end{figure}

\begin{figure*}
    \includegraphics[width=\textwidth]{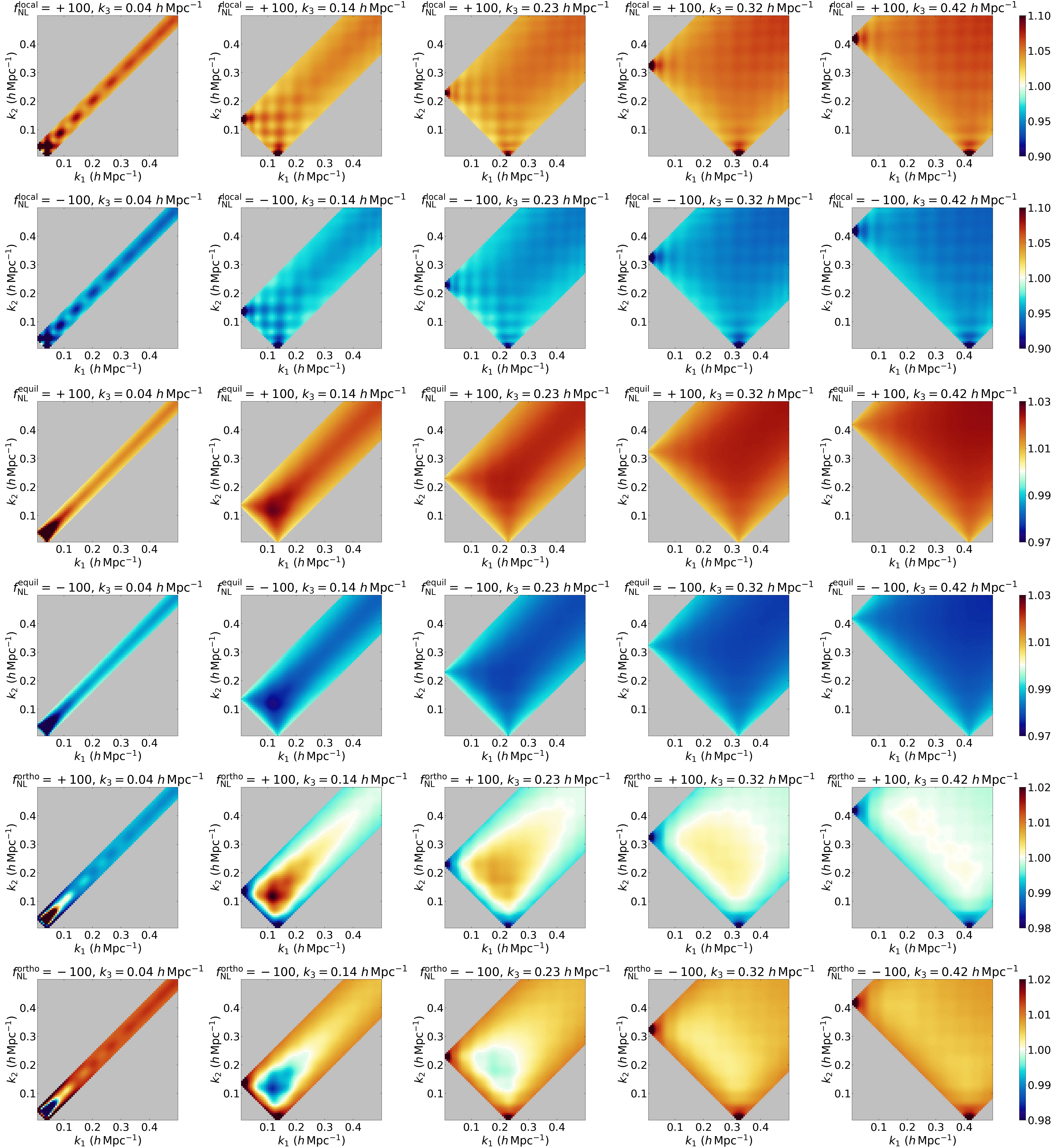}
    \caption{The impact of PNG on the matter bispectrum at $z=1$ (ratio NG/G, averaged from $500$ N-body simulations). We show the local (top rows), equilateral (middle rows) and orthogonal (bottom rows) shapes.}
    \label{fig:bisp-png}
\end{figure*}

As an initial visual check, in figure~\ref{fig:pk-png} we show the impact on the power spectrum of including PNG in the simulations at redshift $z=1$. As expected, such impact is essentially negligible on linear scales and rises to percent level, at most, in the non-linear regime, despite the large NG amplitudes in input ($\fNL=\pm100$). Moreover, the effect of local and equilateral NG is very degenerate. In figure~\ref{fig:bisp-png}, we show the same comparison for bispectrum configurations, reconstructed from our modal basis, by using eq.\ \ref{eq:modal-bispectrum}, where the details in calculating the $\gamma$ matrix, needed for this step, can be found in appendix \ref{app:gamma}. While the effect remains at percent level for a single triangle, we have now of course a much larger number of available configurations. Moreover, we can clearly see how the effect of the local primordial template peaks in the squeezed limit -- where one $k$ is much smaller than the other two -- whereas the equilateral shape produces the largest signal when three $k$ are of the same order. The orthogonal bispectrum has both squeezed and equilateral components. 

\begin{figure}
    \includegraphics[width=0.99\linewidth]{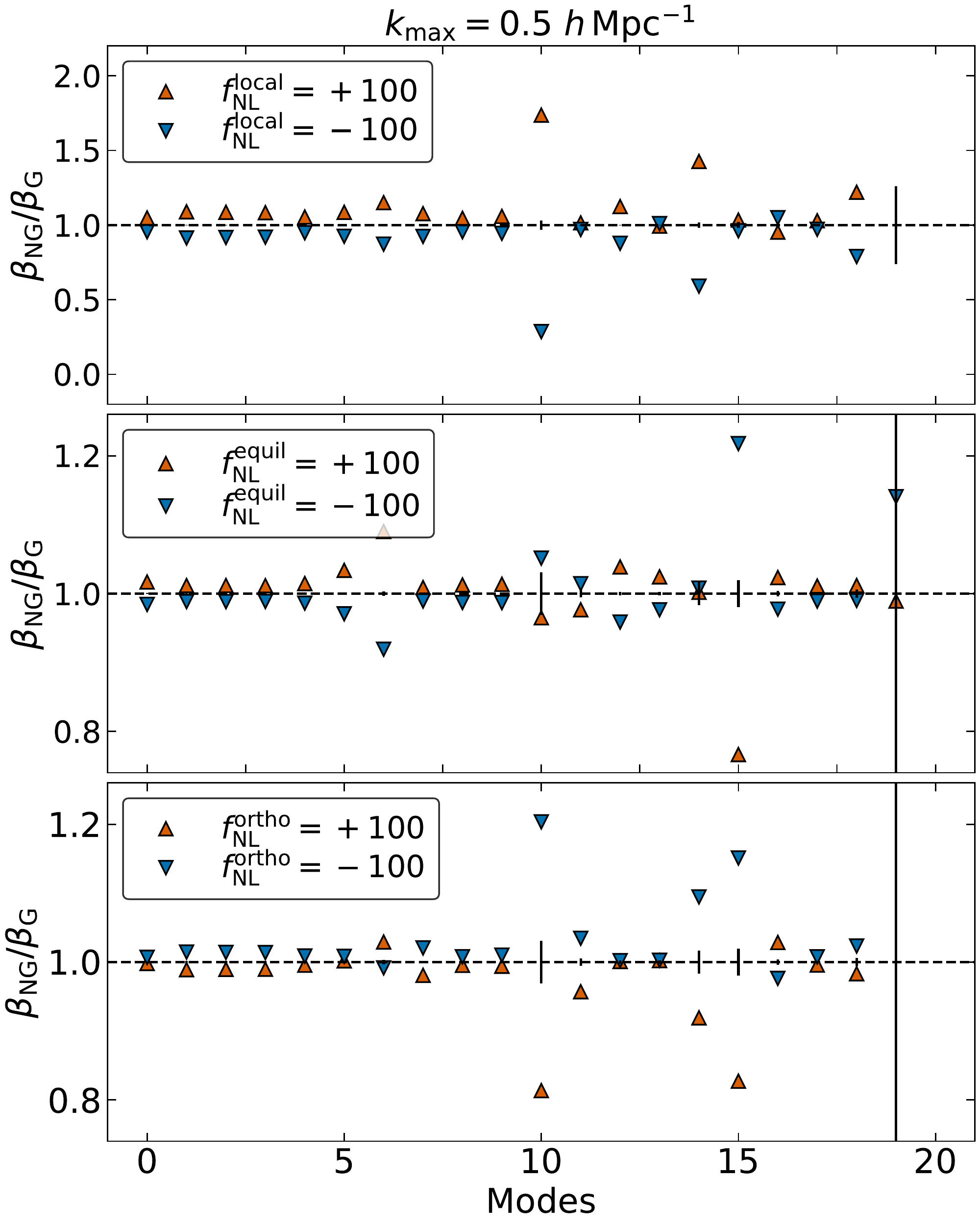}
    \caption{The impact of PNG on the modal bispectrum coefficients (ratio NG/G, averaged from $500$ N-body simulations) at $z=1$. The vertical black lines show the standard errors for 500 Gaussian fiducial simulations, for most modes these are unobservably small.}
    \label{fig:modes-png}
\end{figure}

\begin{figure*}
    \includegraphics[width=0.33\linewidth]{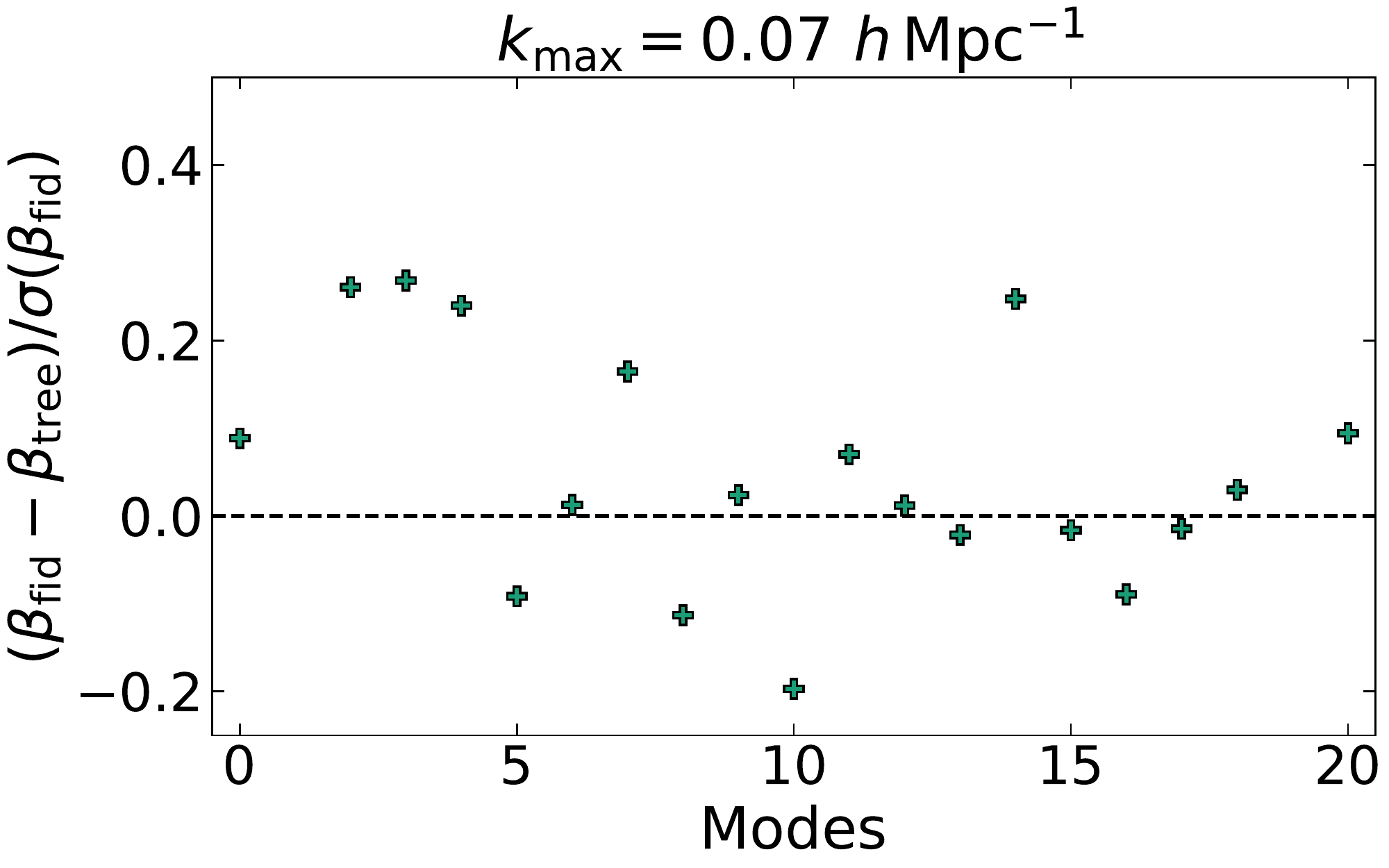}
    \includegraphics[width=0.33\linewidth]{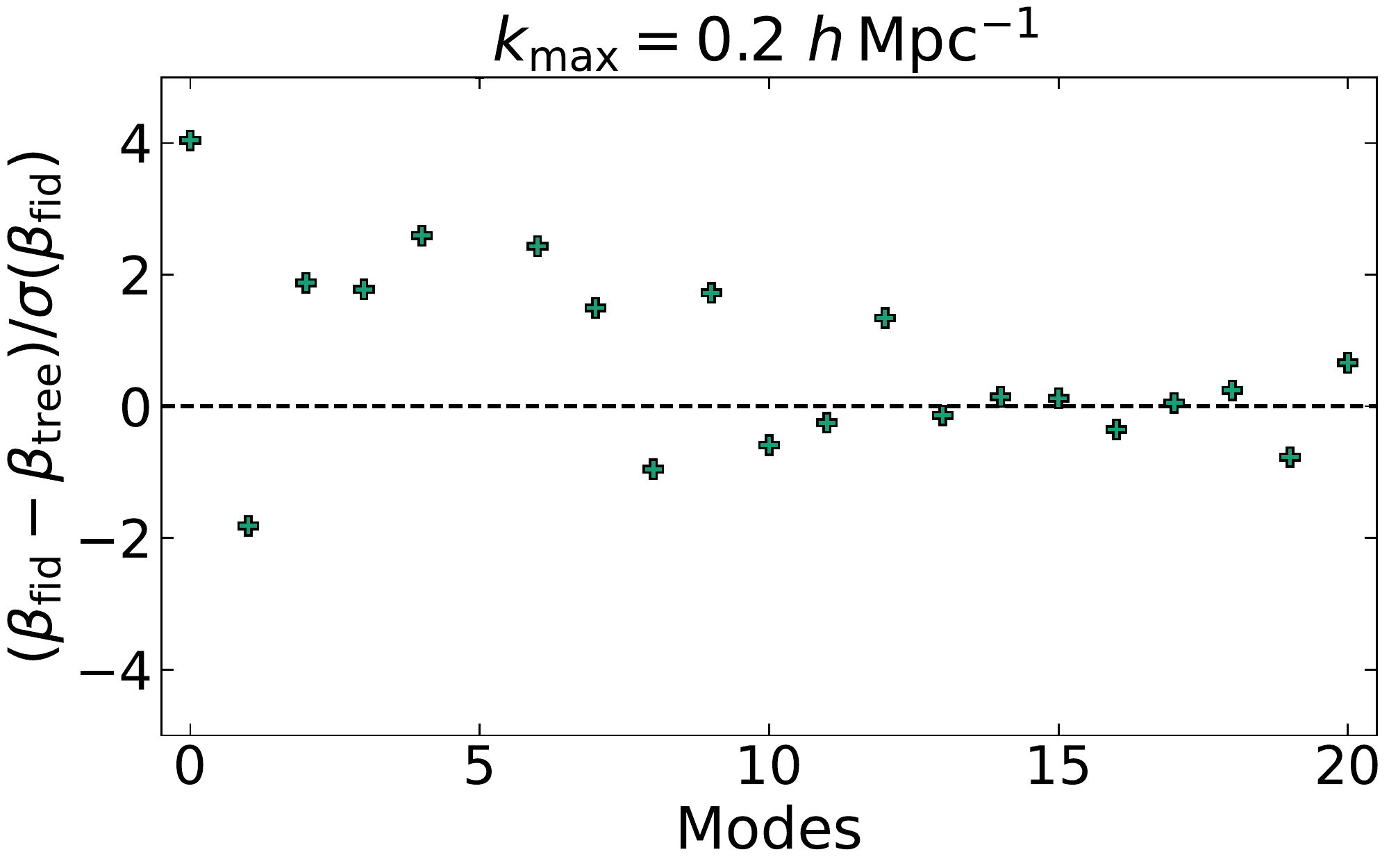}
    \includegraphics[width=0.33\linewidth]{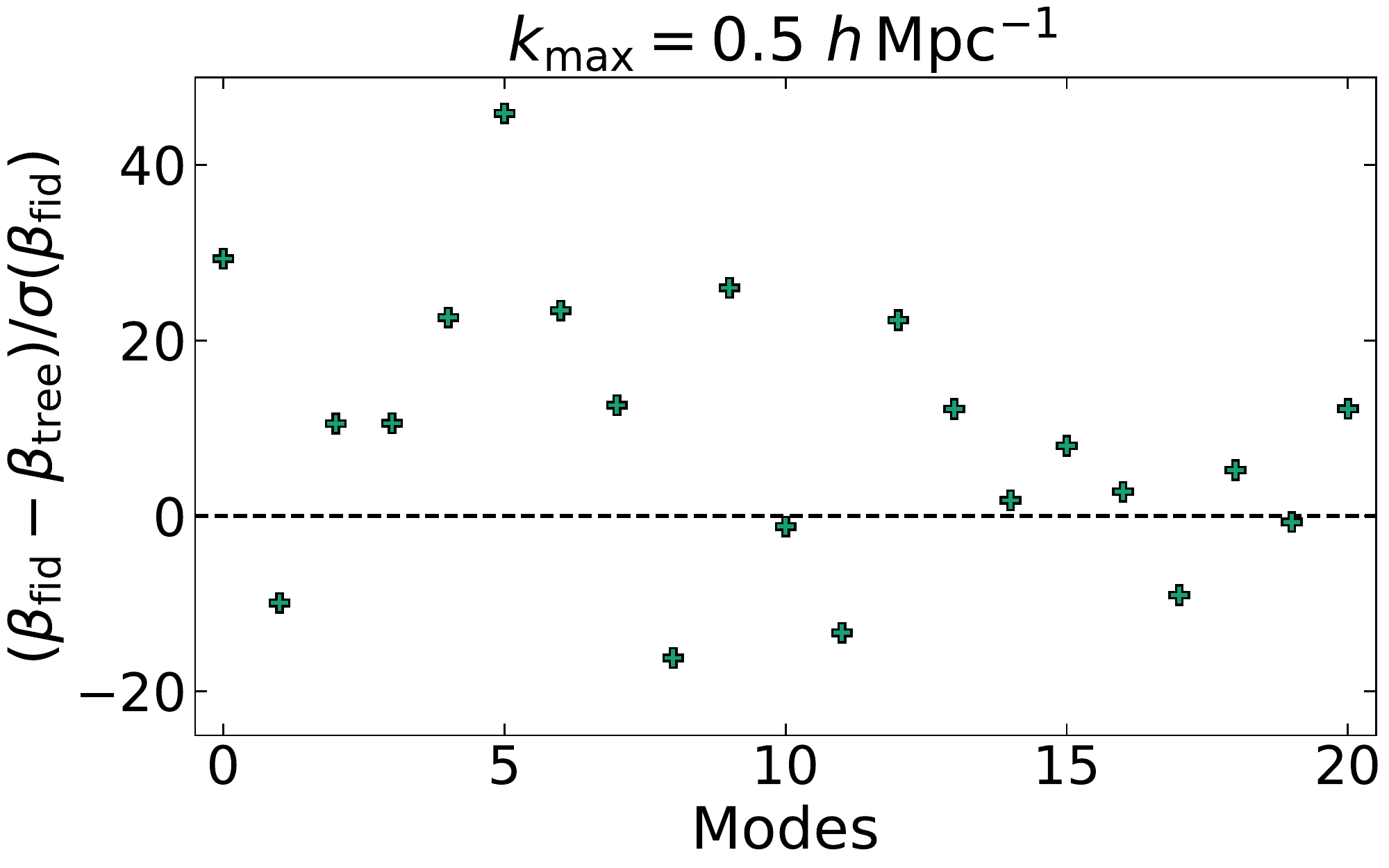}
    \caption{The difference between the modal bispectrum coefficients of the \Quijote~N-body simulations (average from $8000$ simulations at fiducial cosmology) and their theoretical prediction computed at tree-level, divided by their respective standard deviation. Note the different vertical scales of each plot.}
    \label{fig:modes-tree}
\end{figure*}

In figures~\ref{fig:modes-png} and \ref{fig:modes-tree}, we show some modal bispectrum coefficients $\beta_n$ of the \Quijote~simulations, evaluated using eq.\ \eqref{eq:beta}. Figure~\ref{fig:modes-png} is the direct equivalent of figure\ \ref{fig:bisp-png} at the modal level, and also highlights the very distinct behaviours of the three primordial shapes. In figure~\ref{fig:modes-tree}, we compare the modal coefficients of the fiducial Gaussian simulations to their theoretical prediction, computed at tree-level. The first five modes correspond to the special functions based on the tree-level and local separable templates (see appendix~\ref{app:gamma} for more details), while the rest are polynomials of increasing degree, from left to right. As expected, the difference between measured and predicted signals increases significantly with $\kmax$. Note also that in this analysis we work with the $Q_n$ separable modes defined in eq.\ \eqref{eq:modal-bispectrum}. These, by definition, are not orthogonal with respect to the inner product defined in eq.\ \eqref{eq:inner} and always display non-trivial correlations. For this reason, in modal bispectrum analyses a new basis is often introduced, by orthogonalizing the $Q_n$ templates. This procedure, however, only removes the correlations in the weakly non-Gaussian case and, after verification, does not bring any improvement to the results presented in this paper. On the contrary it can add some numerical instability when including high-order polynomial modes; thus, it is not included here.

\begin{figure*}
    \includegraphics[width=0.99\textwidth]{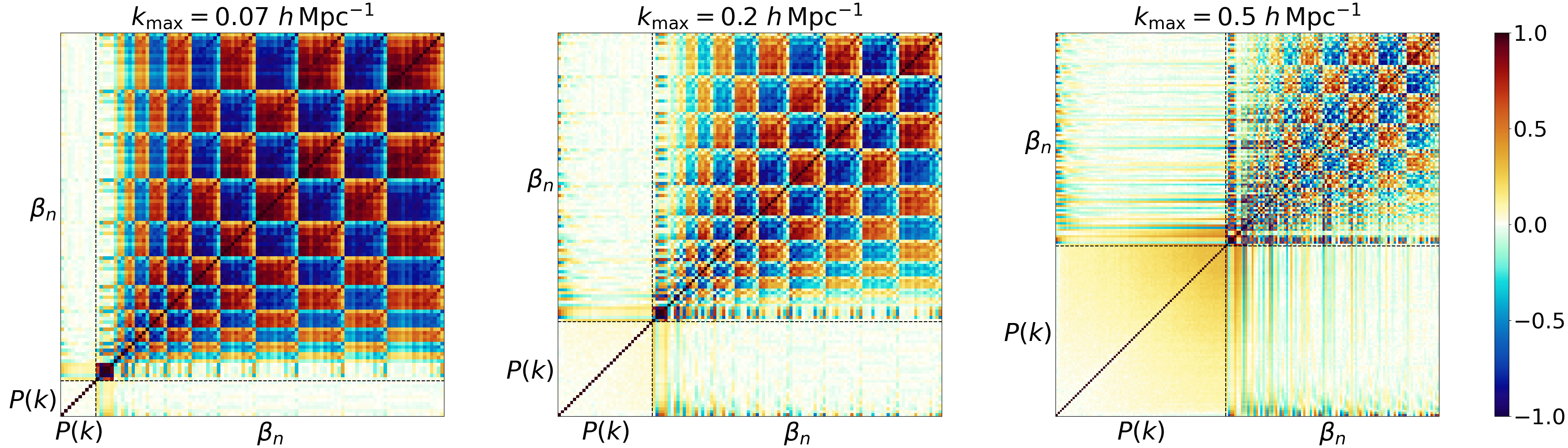}
    \caption{The correlation matrix of the power spectrum $P(k)$ and modal bispectrum coefficients $\beta_n$ of the \Quijote~simulations for different $\kmax$. We include a similar number of bispectrum modes in each case, using polynomial basis functions up to a degree of 12.}
    \label{fig:correlations}
\end{figure*}

After extracting summary statistics from the data and checking that they behave according to expectation, the next step of the analysis is the optimal compression to the score function, following the procedure described in section~\ref{sec:compression}. This requires the evaluation of the mean and the inverse covariance matrix of the estimated power spectrum bins and bispectrum modal coefficients, as well as their derivatives with respect to each considered parameter (cosmological + NG amplitudes). We compute the data covariance $\hat{\cov}$ from $8000$ \Quijote~N-body simulations at fiducial cosmology and we apply the Hartlap/Anderson correction factor \citep{Hartlap:2006kj}, to obtain an unbiased estimate of the precision matrix:
\begin{equation}
    \cov^{-1} = \frac{n_\mathrm{r}-n_\mathrm{d}-2}{n_\mathrm{r}-1}\hat{\cov}^{-1},
    \label{eq:hartlap}
\end{equation}
where $n_\mathrm{r}$ is the number of realizations ($8000$ here) and $n_\mathrm{d}$ represents the size of the data vector (a few hundreds at most here).\footnote{This correction factor is never more than a few $\%$ here, a regime where it has been verified \citep[see e.g.][]{Gualdi:2021yvq, Gil-Marin:2022hnv} that it gives results similar to the more advanced method developed in \citet{2016MNRAS.456L.132S}, when estimating Fisher bounds.} In figure~\ref{fig:correlations}, we show the full correlation matrix given by $\cov_{ij}/\sqrt{\cov_{ii}\cov_{jj}}$, including both power spectrum bins $P(k)$ and mode amplitudes $\beta_n$. As expected, the power spectrum becomes more correlated on smaller scales. Bispectrum modal coefficients however have the opposite behaviour, becoming less correlated at higher $\kmax$. This is due to the fact that each mode is, by definition, the fit of a given template (polynomial, or simple separable bispectrum shape) to the data, including all scales in the range $[\kmin,\kmax]$. Thus, an increased $\kmax$ helps to better distinguish between them, leading to lower correlations. Note also the non-zero correlation between the power spectrum and the bispectrum, which will play an important role in the analyses presented in this paper. We do not include super-sample covariance terms in our analyses, which have a negligible impact on our results as shown in our companion paper \citep{Coulton:2022}. We also use these $8000$ simulations to compute the mean value of the modal coefficient vector, which fixes the score function to zero at fiducial cosmology with no PNG. 

The derivatives can be calculated quickly from the sets of $500$ simulations with one adjusted parameter and matching random seeds,\footnote{In addition of using matching random seeds, different step values between simulations have been tested in order to reduce the numerical noise when evaluating the derivatives.} using the standard formula:
\begin{equation}
    \frac{\partial \mathrm{d}_n}{\partial\theta}\simeq\frac{\mathrm{d}_n(\theta_\mathrm{fid} + \Delta\theta) - \mathrm{d}_n(\theta_\mathrm{fid} - \Delta\theta)}{2\Delta\theta},
    \label{eq:derivative}
\end{equation}
where $\theta$ can be any of the cosmological or NG amplitude parameters considered in this paper and $\mathrm{d}$ are either power spectrum or bispectrum mode estimates. 

\subsection{Fisher constraints}
\label{sec:constraints}

With the numerically evaluated, joint covariance matrix of power spectrum bins and bispectrum modes, described in the previous section, we can derive expected error bars for the quasi-maximum likelihood estimator (eq.\ \ref{eq:ml-estimator}), using the Fisher matrix defined in eq.\ \eqref{eq:fisher}. 

In table~\ref{tab:constraints}, we present the resulting constraints on cosmological parameters $\sigma_8$, $\Omega_m$,  $\Omega_b$, $n_s$, $h$ and PNG amplitude parameters $\fNLloc$, $\fNLeq$, $\fNLort$, for the three different $\kmax$ studied.  We consider the power spectrum and the bispectrum (modal coefficients) separately and then analyze them jointly, in order to gain insight on their relative constraining power at different scales. 

In the upper part of the table, we assume Gaussian initial conditions. As pointed out in \citet{Hahn:2019zob}, we see in this case that adding bispectrum information does help to improve power spectrum constraints on cosmological parameters. While the gain is marginal on linear scales, it becomes significant in the non-linear regime, where the bispectrum alone is actually a more powerful probe than the power spectrum alone, for all considered parameters, with the exception of $\sigma_8$. 

In the lower part of the table, we include both cosmological parameters and PNG of the three standard shapes. In this case, if we look at bounds from the power spectrum alone, we not only see that PNG amplitudes are essentially unconstrained, as expected, but also that errors on other parameters significantly degrade in comparison with the primordial Gaussian case. This is due to degeneracies arising between $\fNL$ and cosmological parameters, which are even more pronounced at non-linear scales. The most affected parameter is $\sigma_8$, the error of which increases by more than a factor $3$. The bispectrum now plays a crucial role not only in constraining PNG at a significant level, but also in breaking such degeneracies and allowing for large improvements in predicted errors for cosmological parameters.  

Note also that in table~\ref{tab:constraints}, we assume that all PNG shapes are simultaneously present in the data. However, we verified that similar degeneracies are introduced even when we consider one NG shape at a time. This is not surprising, remembering that introducing NG of different types produces a similar impact on the power spectrum (see figure~\ref{fig:pk-png}).

When combining the power spectrum and bispectrum information, we recover very similar error bars on cosmological parameters as in the purely Gaussian case, meaning that degeneracies are strongly broken. For the same reason, we also see that PNG bounds from the bispectrum alone are improved by a factor $\sim 2$ when we add power spectrum measurements, despite the fact that the power spectrum yields almost no PNG constraining power by itself. Interestingly, even if we completely fix cosmological parameters in the analysis, combining power spectrum and bispectrum still produces some improvement in PNG constraints. In this case, degeneracies clearly cannot play a role. It turns out that the improvements are now due to correlations between the power spectrum and the bispectrum, displayed in figure \ref{fig:correlations} (intuitively, at tree level, the gravitational bispectrum is linked to the power spectrum squared, via a convolution kernel; hence, measuring the latter produces an information gain also on the former. A better knowledge of the gravitational bispectrum allows in turn for a better separation of this component from the primordial signal).\footnote{An additional role is played by PNG contributions to the power spectrum, at loop-level, if we perform a single shape analysis. Such contributions are however degenerate among different shapes.}  

In table~\ref{tab:constraints-independent}, we report the independent Fisher bounds for the three PNG shapes, after marginalizing over cosmological parameters.

\begin{figure*}
    \includegraphics[width=0.99\textwidth]{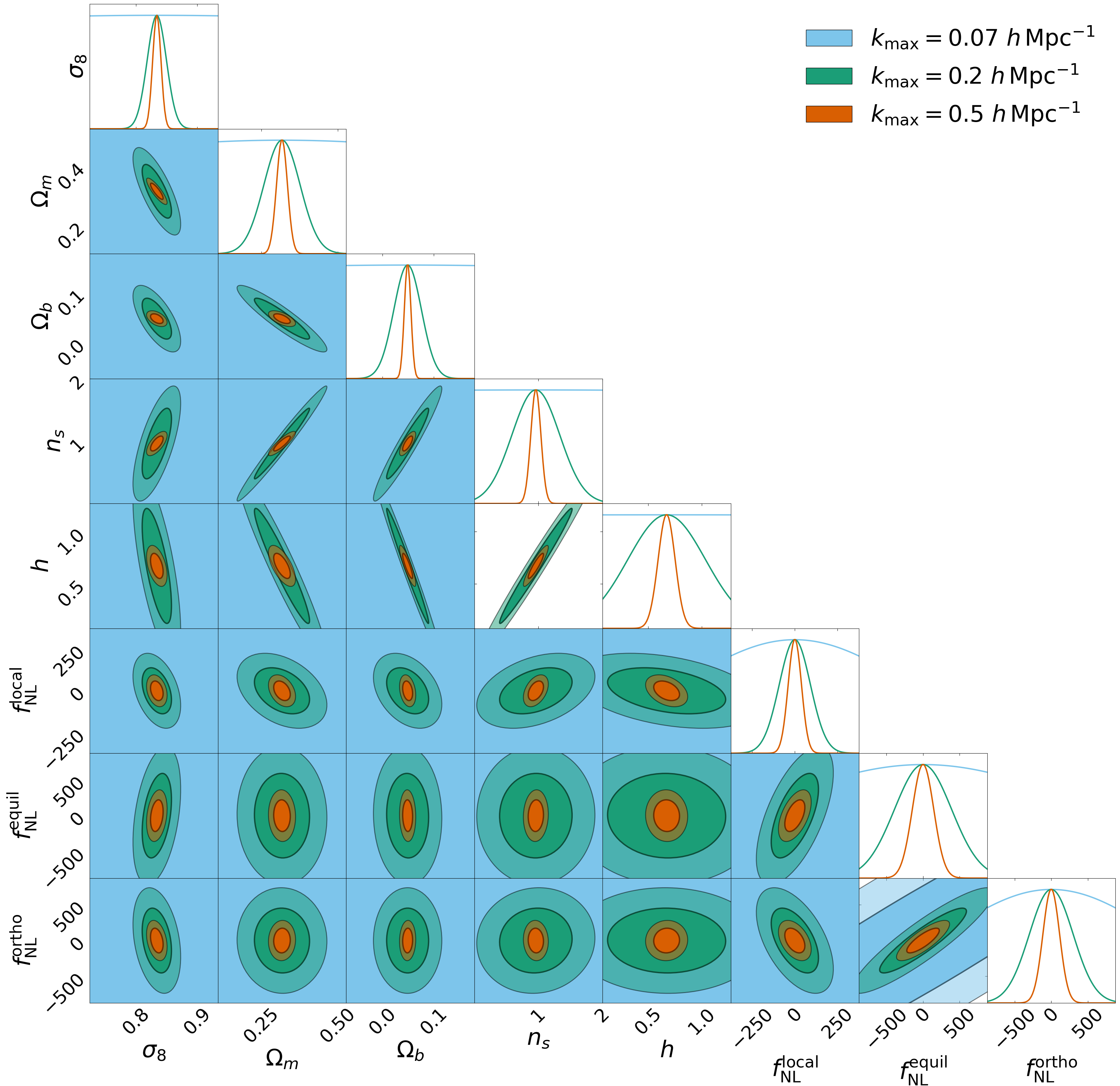}
    \caption{Joint constraints on cosmological parameters and PNG from the power spectrum and the modal bispectrum for different $\kmax$ at $z=1$.}
    \label{fig:contours}
\end{figure*}

Another important result is that probing non-linear scales significantly improves the constraints on every considered parameter. This is further illustrated in figure~\ref{fig:contours}, where we show the contours obtained using jointly the power spectrum and the bispectrum for different $\kmax$. A more detailed analysis of the information content of different scales is in our companion paper \citep{Coulton:2022}, where we show that errors tend to saturate at $\kmax=0.3~\hMpc$, at $z=0$, due to correlations between Fourier modes at strongly non-linear scales.

\begin{deluxetable*}{c|ccccccccc}
 \tablecaption{Joint constraints on cosmological parameters and PNG from the power spectrum and the modal bispectrum at $z=1$, for different $\kmax$. We analyzed $8000$ \Quijote~N-body simulations of $1~(\mathrm{Gpc}/h)^3$ volume at fiducial cosmology, and sets of $500$ N-body simulations with one adjusted parameter.  \label{tab:constraints}}
\tablehead{        & $\kmax$ & $\sigma_8$ & $\Omega_m$ & $\Omega_b$ & $n_s$ & $h$ & $\fNLloc$ & $\fNLeq$ & $\fNLort$\\
        & ({\footnotesize $\hMpc$})  & 0.834 & 0.3175 & 0.049 & 0.9624 & 0.6711 & 0 & 0 & 0}
\startdata
\hline \hline
$P(k)$  
& \textbf{0.07} & $\pm 0.17$ & $\pm 0.32$ & $\pm 0.32$ & $\pm 3.9$ & $\pm 4.4$ & & & \\
& \textbf{0.2} & $\pm 0.012$ & $\pm 0.039$ & $\pm 0.018$ & $\pm 0.24$ & $\pm 0.24$ & & & \\
& \textbf{0.5} & $\pm 0.0045$ & $\pm 0.011$ & $\pm 0.0062$ & $\pm 0.042$ & $\pm 0.062$ & & & \\
\hline
$\beta_n$  
& \textbf{0.07} & $\pm 0.59$ & $\pm 0.95$ & $\pm 1.$ & $\pm 12$ & $\pm 14$ & & & \\
& \textbf{0.2} & $\pm 0.014$ & $\pm 0.051$ & $\pm 0.023$ & $\pm 0.32$ & $\pm 0.31$ & & & \\
& \textbf{0.5} & $\pm 0.0063$ & $\pm 0.016$ & $\pm 0.006$ & $\pm 0.066$ & $\pm 0.071$ & & & \\
\hline
$P(k) + \beta_n $  
& \textbf{0.07} &  $\pm 0.17$ & $\pm 0.31$ & $\pm 0.31$ & $\pm 3.7$ & $\pm 4.2$ & & & \\
& \textbf{0.2} & $\pm 0.011$ & $\pm 0.035$ & $\pm 0.015$ & $\pm 0.21$ & $\pm 0.21$ & & & \\
& \textbf{0.5} & $\pm 0.0038$ & $\pm 0.0091$ & $\pm 0.0048$ & $\pm 0.031$ & $\pm 0.048$ & & & \\
\hline\hline
$P(k)$  
& \textbf{0.07} & $\pm 0.55$ & $\pm 0.41$ & $\pm 0.36$ & $\pm 4.5$ & $\pm 4.9$ & $\pm 200000$ & $\pm 500000$ & $\pm 180000$ \\
& \textbf{0.2} & $\pm 0.13$ & $\pm 0.07$ & $\pm 0.031$ & $\pm 0.49$ & $\pm 0.39$ & $\pm 31000$ & $\pm 85000$ & $\pm 36000$ \\
& \textbf{0.5} & $\pm 0.069$ & $\pm 0.035$ & $\pm 0.013$ & $\pm 0.24$ & $\pm 0.18$ & $\pm 9600$ & $\pm 29000$ & $\pm 11000$ \\
\hline
$\beta_n$  
& \textbf{0.07} & $\pm 1.1$ & $\pm 1.3$ & $\pm 1.4$ & $\pm 18$ & $\pm 20$ & $\pm 670$ & $\pm 2300$ & $\pm 1500$ \\
& \textbf{0.2} & $\pm 0.016$ & $\pm 0.059$ & $\pm 0.027$ & $\pm 0.37$ & $\pm 0.36$ & $\pm 91$ & $\pm 390$ & $\pm 300$ \\
& \textbf{0.5} & $\pm 0.0068$ & $\pm 0.018$ & $\pm 0.0064$ & $\pm 0.078$ & $\pm 0.079$ & $\pm 39$ & $\pm 150$ & $\pm 110$ \\
\hline
$P(k) + \beta_n $  
& \textbf{0.07} & $\pm 0.17$ & $\pm 0.31$ & $\pm 0.31$ & $\pm 3.7$ & $\pm 4.2$ & $\pm 350$ & $\pm 930$ & $\pm 610$ \\
& \textbf{0.2} & $\pm 0.011$ & $\pm 0.035$ & $\pm 0.015$ & $\pm 0.21$ & $\pm 0.21$ & $\pm 52$ & $\pm 170$ & $\pm 120$ \\
& \textbf{0.5} & $\pm 0.0038$ & $\pm 0.0093$ & $\pm 0.0048$ & $\pm 0.033$ & $\pm 0.048$ & $\pm 22$ & $\pm 94$ & $\pm 58$ \\
    \hline
\enddata
\end{deluxetable*}

\begin{table}
\begin{center} 
    {\begin{tabular}{ccc}
        $\fNLloc$ & $\fNLeq$ & $\fNLort$ \\
        \hline\hline
         $\pm 16$ & $\pm 77$ & $\pm 40$ \\
        \hline
    \end{tabular}} 
 \caption{Fisher $1$-$\sigma$ constraints on the 3 PNG shapes from the power spectrum and the modal bispectrum at $\kmax=0.5~\hMpc$, for a cubic volume of $1~(\mathrm{Gpc}/h)^3$ and at $z=1$, after marginalization over cosmological parameters. Each shape is analyzed independently. }
 \label{tab:constraints-independent}
\end{center}
\end{table}

The accuracy of the modal expansion of the bispectrum has a strong impact on expected error bars and thus it is important to verify that the modal results presented in this section are fully converged. There are several solutions to improve the modal reconstruction, in order to achieve higher accuracy with as small as possible number of modes. Some of them involve using custom separable templates, corresponding to, or highly correlated with, expected signals in the data. 
In this work, we already pursue this approach by including the tree-level matter bispectrum and the primordial local shape in the modal basis. In the future, we plan to explore it further, by considering, for example, separable templates that are strongly correlated to the predicted bispectrum at one-loop in perturbation theory, or describing the other primordial shapes.

\begin{figure*}
    \includegraphics[width=0.49\textwidth]{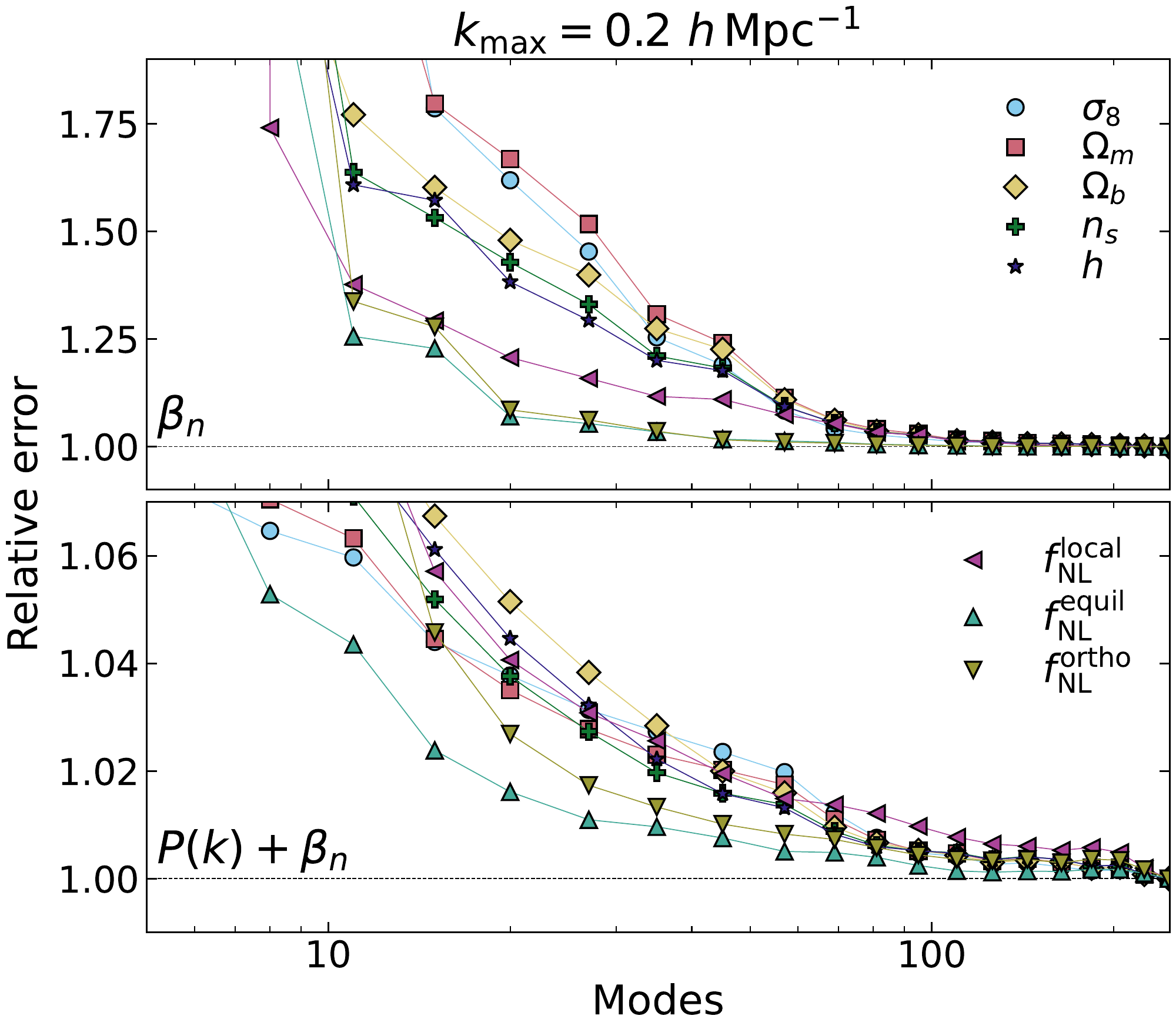}
    \includegraphics[width=0.49\textwidth]{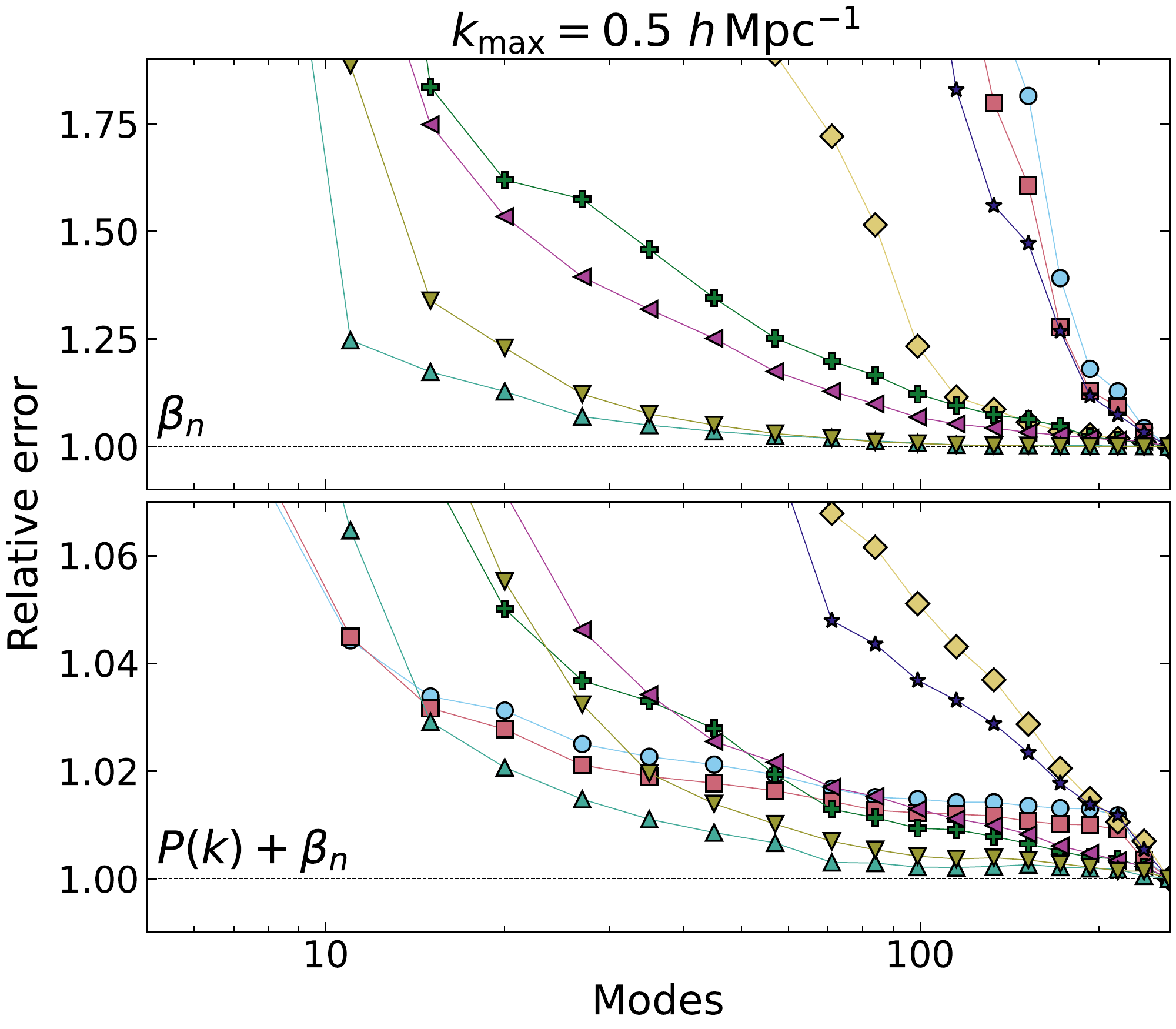}
    \caption{The convergence of constraints on cosmological parameters and PNG as a function of the number of modes used for the modal bispectrum estimation. The top row corresponds to the bispectrum only case, while the bottom row also includes the power spectrum information. Note the difference of scale between the top and bottom rows. For each parameter, error bars are divided by the corresponding minimum, see table~\ref{tab:constraints} for the actual values. From left to right, each data point corresponds to include polynomials of one degree higher in the modal basis.}
    \label{fig:constraints}
\end{figure*}

The obvious, although less economical, alternative approach to improve reconstruction accuracy simply consists of including additional, higher order polynomials in the basis. We consider here polynomials up to the degree $20$, which are combined in triplets to obtain several hundred bispectrum modes. In figure~\ref{fig:constraints}, we study the convergence of the constraints as a function of the number of modes included in the modal expansion, each data point from left to right corresponding to adding polynomials of one degree higher. As could be expected, the accurate description of the bispectrum on small scales requires more modes. The constraints on some cosmological parameters are not even fully converged by taking polynomials up to degree $20$, making our Fisher forecast in such case slightly overly conservative; we verify that the results on primordial non-Gaussianity are however always converged. The situation is significantly better with the joint power spectrum and bispectrum analysis. In the mildly non-linear regime, we achieve convergence with $\sim 100$ modes, and using only $10$ modes produces constraints that are already within $10\%$ of their converged values. Even in the non-linear regime, going beyond $\sim 100$ modes we improve the Fisher bounds only by $\sim 1\%$ (except for $\Omega_b$ and $h$ for which it is of a few percent). This means that, for the combined power spectrum and bispectrum analysis, we can stop the polynomial expansion to a lower degree than $20$, with a negligible loss of information. To be exact, the constraints given in table~\ref{tab:constraints} use polynomials up to the degree $12$ for the joint power spectrum $+$ bispectrum case, and $20$ for the bispectrum only analysis. Note that using higher order polynomials can add numerical noise to the analysis, thus stopping the expansion at a reasonable degree, like we do here, is a more robust approach.  

The numbers discussed above show that the modal representation is quite efficient. Compare, for example, the $\sim 100$ modal coefficients in the current analysis to the thousands of $k$-triplets that are typically necessary to study the bispectrum at the same non-linear scales, using the standard, ``binned'' estimator (e.g.\ $\sim 2000$ triangle configurations with rather large bins of width $3k_F$). This more efficient compression is, for example, important to increase the accuracy of the numerical derivatives evaluated from simulations. 

\begin{figure*}
    \includegraphics[width=0.49\linewidth]{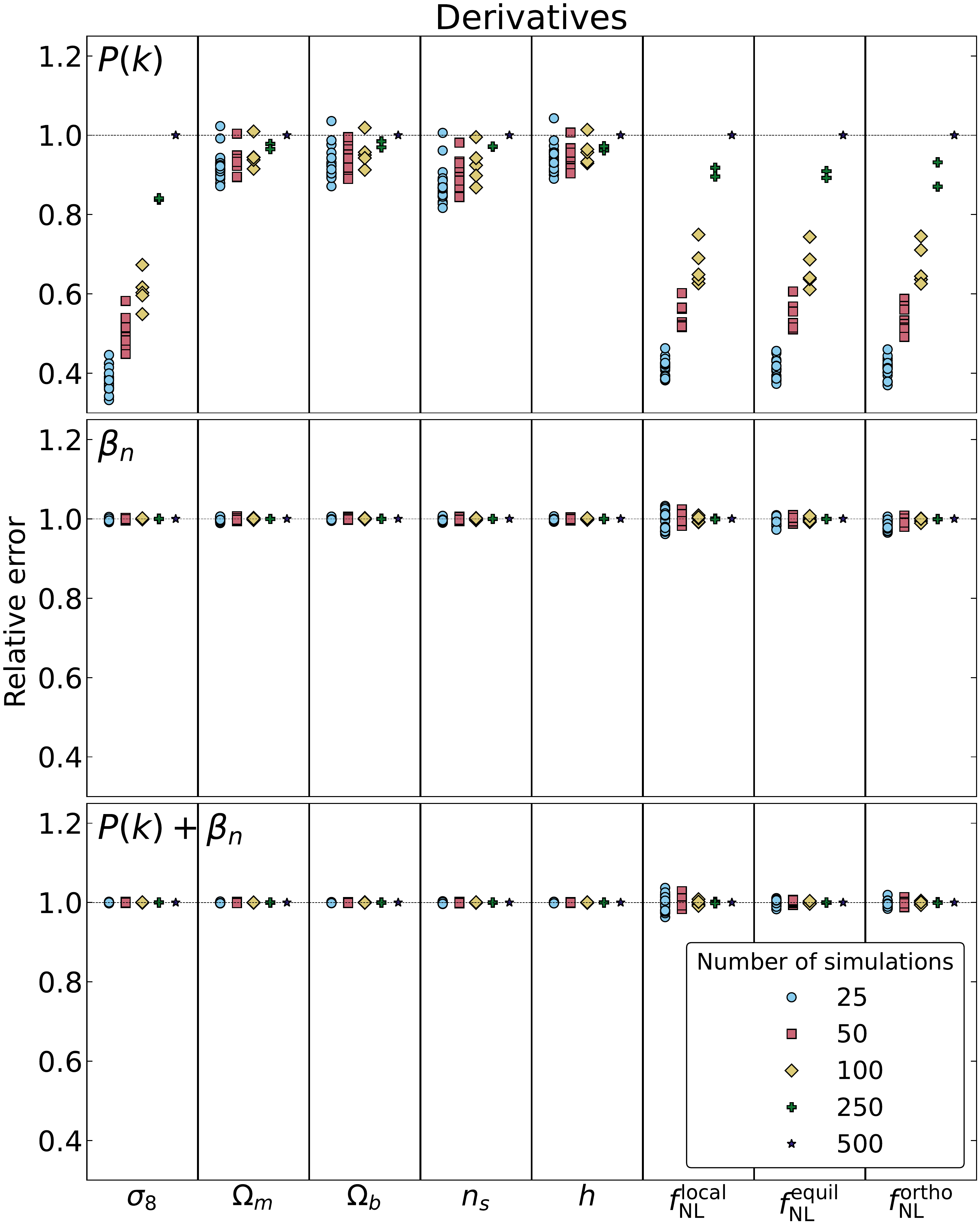}
    \includegraphics[width=0.49\linewidth]{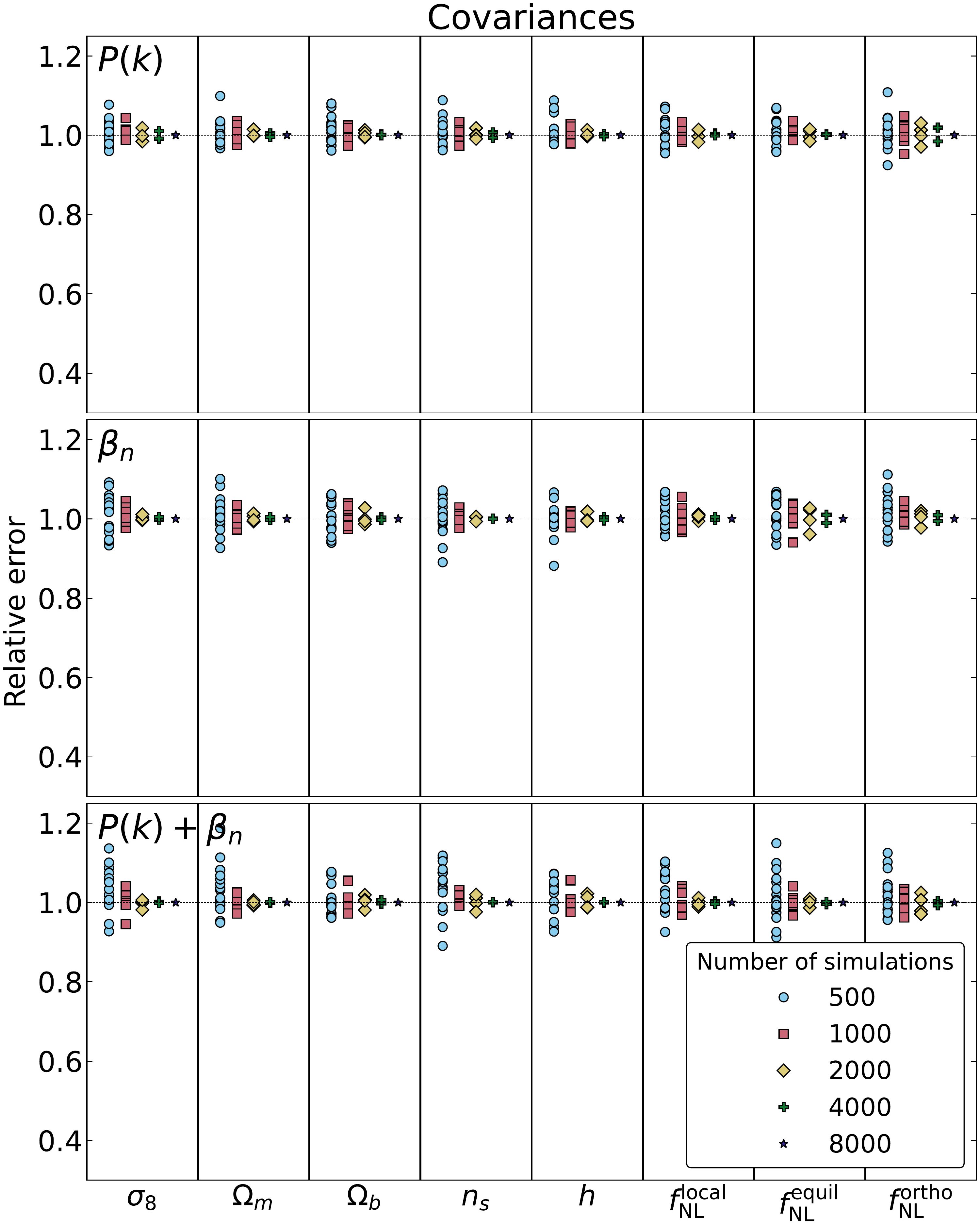}
    \caption{The stability of Fisher constraints under variations of the number of simulations. In the left panels, we vary the number of simulations used to calculate numerical derivatives, while on the right panels we vary the number of simulations used to measure covariances. All error bars are normalized by the result obtained using the full dataset, see table~\ref{fig:constraints} for actual values. The top, middle and bottom rows correspond respectively to power spectrum only, bispectrum only and combined power spectrum and bispectrum analyses. Each column corresponds to the constraints on one parameter, and they are all analyzed jointly. We consider $5$ different numbers of simulations to compute derivatives (from $25$ to $500$) and covariances (from $500$ to $8000$), each number corresponding to its own colour and marker. We include scales down to $\kmax=0.5~\hMpc$.}
    \label{fig:convergence-fisher}
\end{figure*}

Convergence tests have to be performed not only for the modal basis, but also with respect to the total number of mock realizations used to evaluate the Fisher matrix. In figure~\ref{fig:convergence-fisher}, we vary the number of simulations used to estimate derivatives and verify the impact of this change on the final forecast. We find that the constraints from the bispectrum and the joint power spectrum $+$ bispectrum analyses remain very stable, even by using $25$ simulations instead of the usual $500$. At the same time, though, we see that the power spectrum-only forecast is not fully converged, even when we use the full $500$ simulation sample,
leading to somewhat over-optimistic constraints in table~\ref{tab:constraints}, when cosmological parameters and PNG are studied jointly. As the power spectrum-only case was already reported as essentially unconstraining for PNG, this does not change any of the main conclusions presented above \citep[see the companion paper][for a more detailed study of this issue, which can for example be mitigated by assuming some prior on $\fNL$]{Coulton:2022}. 

In figure~\ref{fig:convergence-fisher}, we also verify that the constraints are highly stable when we change the number of simulations (at fiducial cosmology) used to estimate covariance matrices. Using only $1000$ instead of $8000$ simulations only leads to variations at the percent level. However, the requirement on covariance precision will become stricter in the next section, where we discuss actual parameter inference.

\subsection{Parameter estimation}
\label{sec:estimation}

After deriving Fisher matrix forecasts, we now build the quasi-maximum likelihood estimator defined in eq.\ \eqref{eq:ml-estimator} and apply it to different datasets, to verify its efficiency to extract accurate cosmological information.

Our first validation test consists of estimating jointly cosmological parameters and PNG amplitudes ($\sigma_8$, $\Omega_m$, $\Omega_b$, $n_s$, $h$, $\fNLloc$, $\fNLeq$, $\fNLort$) in the \Quijote\ simulations, for different cosmologies. We include non-linear scales ($\kmax=0.5~\hMpc$) and use both the power spectrum and the bispectrum in this analysis, as their combination should lead to smaller error bars, as verified in the previous section. 

\begin{figure}
    \includegraphics[width=0.99\linewidth]{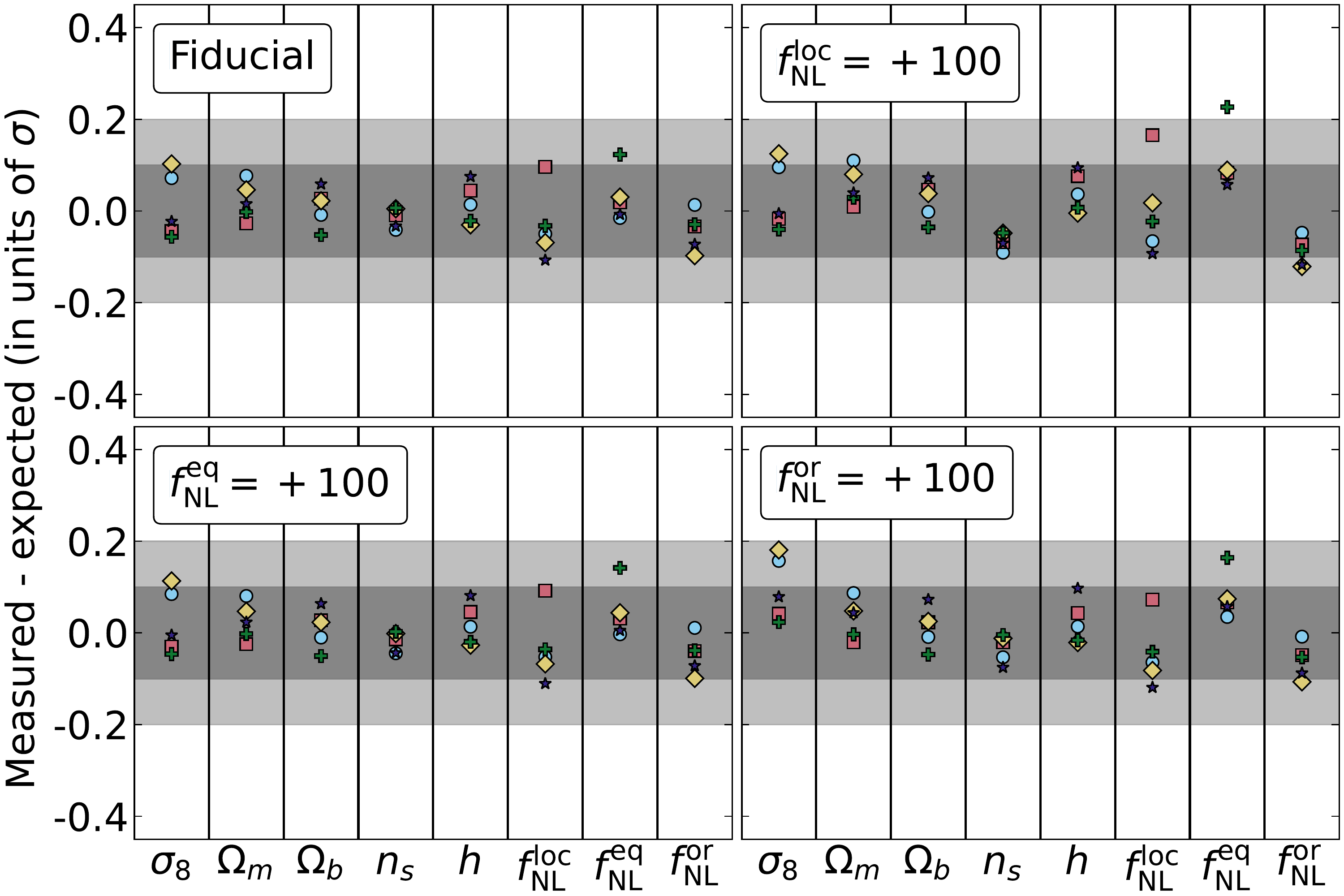}
    \caption{The unbiasedness of the quasi-maximum likelihood estimator (eq.\ \ref{eq:ml-estimator}), when measuring cosmological parameters and PNG amplitudes. We use the power spectrum and the bispectrum jointly, up to $\kmax=0.5~\hMpc$. Each individual column corresponds to a given parameter (cosmological or PNG). To be exact, we show the difference between the expected and measured values, divided by the respective Fisher error bar. Each panel corresponds to a different cosmology of the data samples (i.e. one with Gaussian initial conditions and the three types of PNG). For each of them, we analyze five independent datasets of 100 realizations, each being indicated by its own colour and marker.}
    \label{fig:estimated-quijote}
\end{figure}

In figure~\ref{fig:estimated-quijote}, we show the corresponding results for fiducial cosmology datasets, as well as others having PNG ($\fNLloc=100$, $\fNLeq=100$ or $\fNLort=100$).\footnote{We focus here on changes in $\fNL$, rather than on other cosmological parameters, both because the main focus of this work is on the study of PNG and because, in the available \Quijote\ set, we can access realizations with $\fNLloc=100$, which is more than a $4$-$\sigma$ deviation from the fiducial value of $\fNL=0$. This allows us to test an interesting regime, in which the data we analyze are generated from parameters which are significantly displaced from the model we use to build the estimator.} We compute averages of the estimated parameters and compare them to their input values in the simulations. The correct values are recovered each time, confirming the unbiasedness of the estimator. Note also that we consider only cases in which these averages are extracted from simulations not overlapping with those used to evaluate derivatives and covariances. In this way, we avoid spurious correlations, which could arise from using overlapping sets of data both to calibrate the estimator and to measure parameters.   

\begin{figure}
    \includegraphics[width=0.99\linewidth]{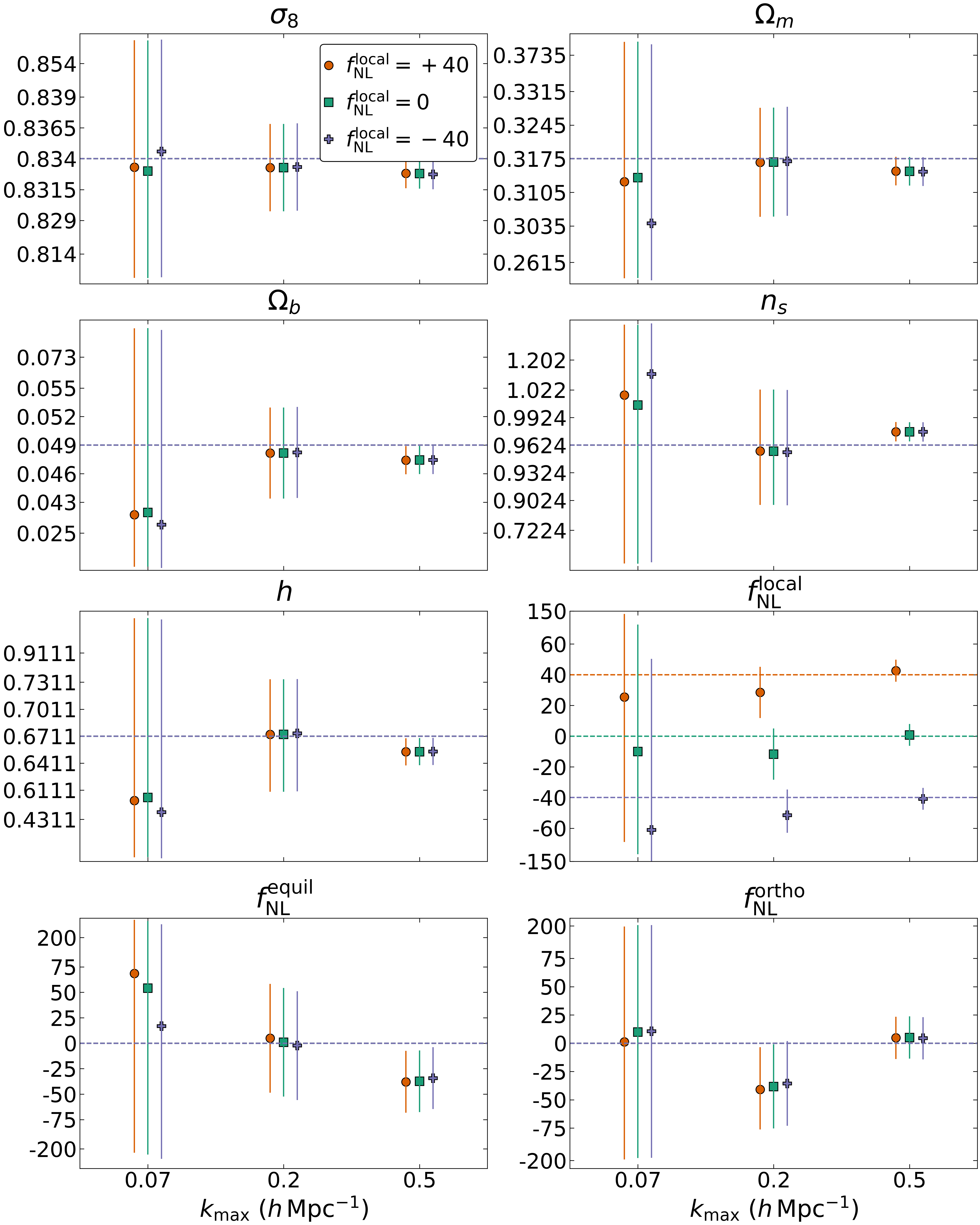}
    \caption{Cosmological and PNG amplitude parameters estimated from several sets of $10$ N-body simulations for different $\kmax$ with a different amount of local PNG in their initial conditions. The orange circles, green squares and purple crosses correspond respectively to $\fNLloc=40$, $0$ and $-40$, and the vertical dashed lines are the expected standard errors on the average from $10$ simulations. The horizontal dashed lines are the expected values of the different datasets. The vertical scale is linear around $0$, and logarithmic close to the edges.}
    \label{fig:estimated-local}
\end{figure}

We also perform a similar analysis for the additional sets of 10 realizations, produced independently of the \Quijote\ simulations and having respectively $\fNLloc=-40$, $0$ and $40$ described in section\ \ref{sec:simulations}. The quasi-maximum likelihood estimator is built from all the available \Quijote\ simulations. Results are shown in figure \ref{fig:estimated-local}, where we focus on the combined power spectrum and modal bispectrum analysis at different $\kmax$. All estimated central values are within or very close to the expected $1$-$\sigma$ range. Moreover, the internal scatter between cosmological parameter estimates for the three datasets ($\fNLloc=-40$, $0$ and $40$) is extremely small. 

In general, we expect the estimator to yield slightly different results, depending on the exact set of simulations used for its construction. For this reason, we perform several stability tests, in which we study variations in the estimated parameters, by changing the number of simulations used to calculate covariances or derivatives. These tests highlight that less accurate derivatives and covariances, which can be obtained if we use a too small number of realizations in their evaluation, have different impact on the final measurements.

\begin{figure}
    \includegraphics[width=0.99\linewidth]{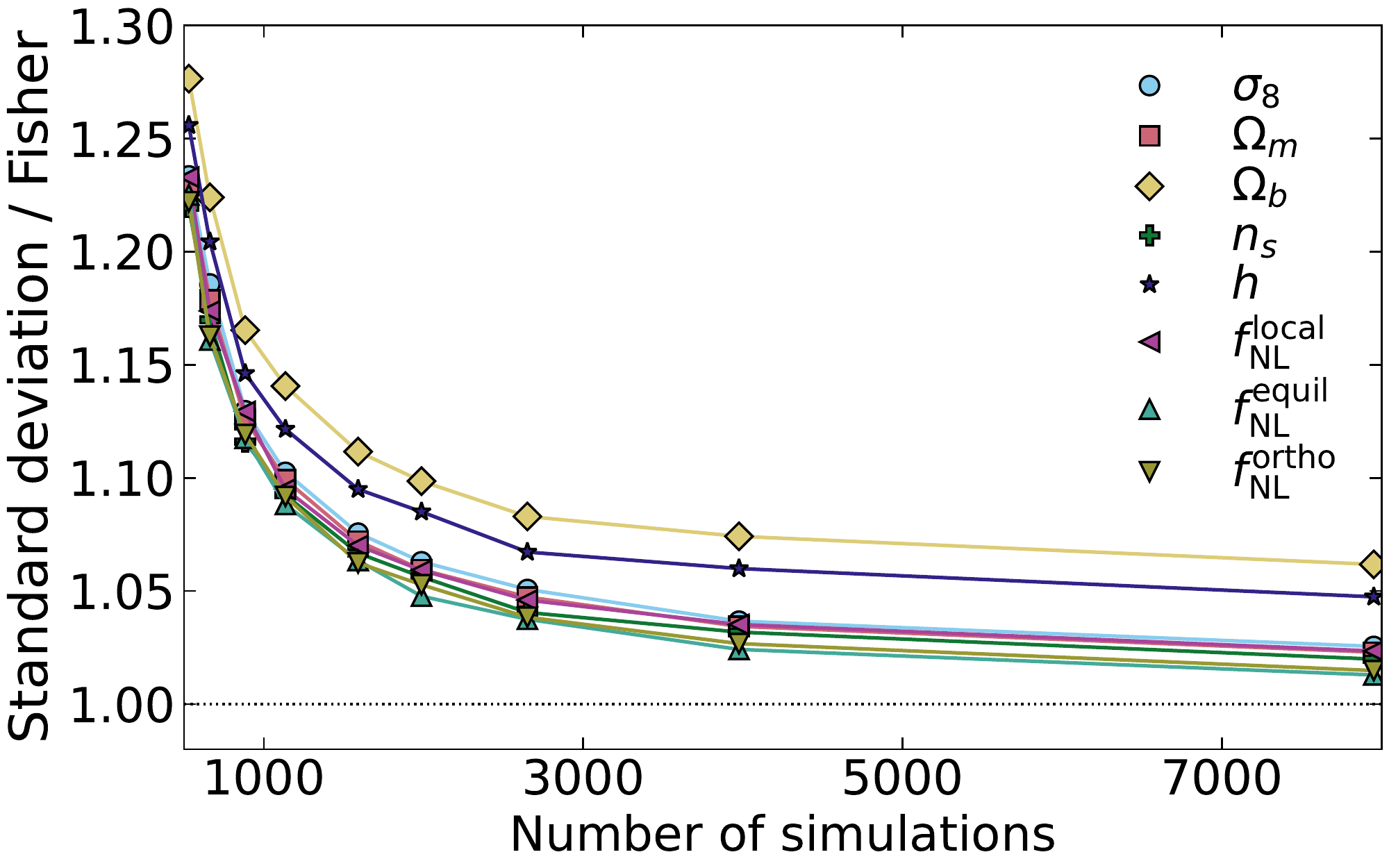}
    \caption{The error bars of the quasi-maximum likelihood estimator as a function of the number of realizations used to calculate the covariance matrix. We analyze the 8000 \Quijote\ simulations at $z=1$, including both the power spectrum and the bispectrum, up to $\kmax=0.5~\hMpc$. All error bars are divided by their corresponding Fisher prediction, as described in section~\ref{sec:constraints}.}
    \label{fig:covariance-estimator}
\end{figure}

A less precise covariance matrix leads to a suboptimal estimation of parameters, as shown in figure~\ref{fig:covariance-estimator}. We compare the error bars of the quasi-maximum likelihood estimator applied to the \Quijote\ simulations at fiducial cosmology, as a function of the number of realizations used to compute covariances, to the Fisher forecasts described in section~\ref{sec:constraints}. This confirms that our estimator yields close to optimal constraints, at the percent level for most parameters ($5\%$ for $\Omega_b$ and $h$), when using many simulations to evaluate the covariance. Using only $1000$ simulations, the increase of error bars is still only of a reasonable order $10\%$, which is small compared to the gain due to including non-linear scales in the analysis (a factor 2 or more for all parameters). 

\begin{figure*}
    \includegraphics[width=0.33\linewidth]{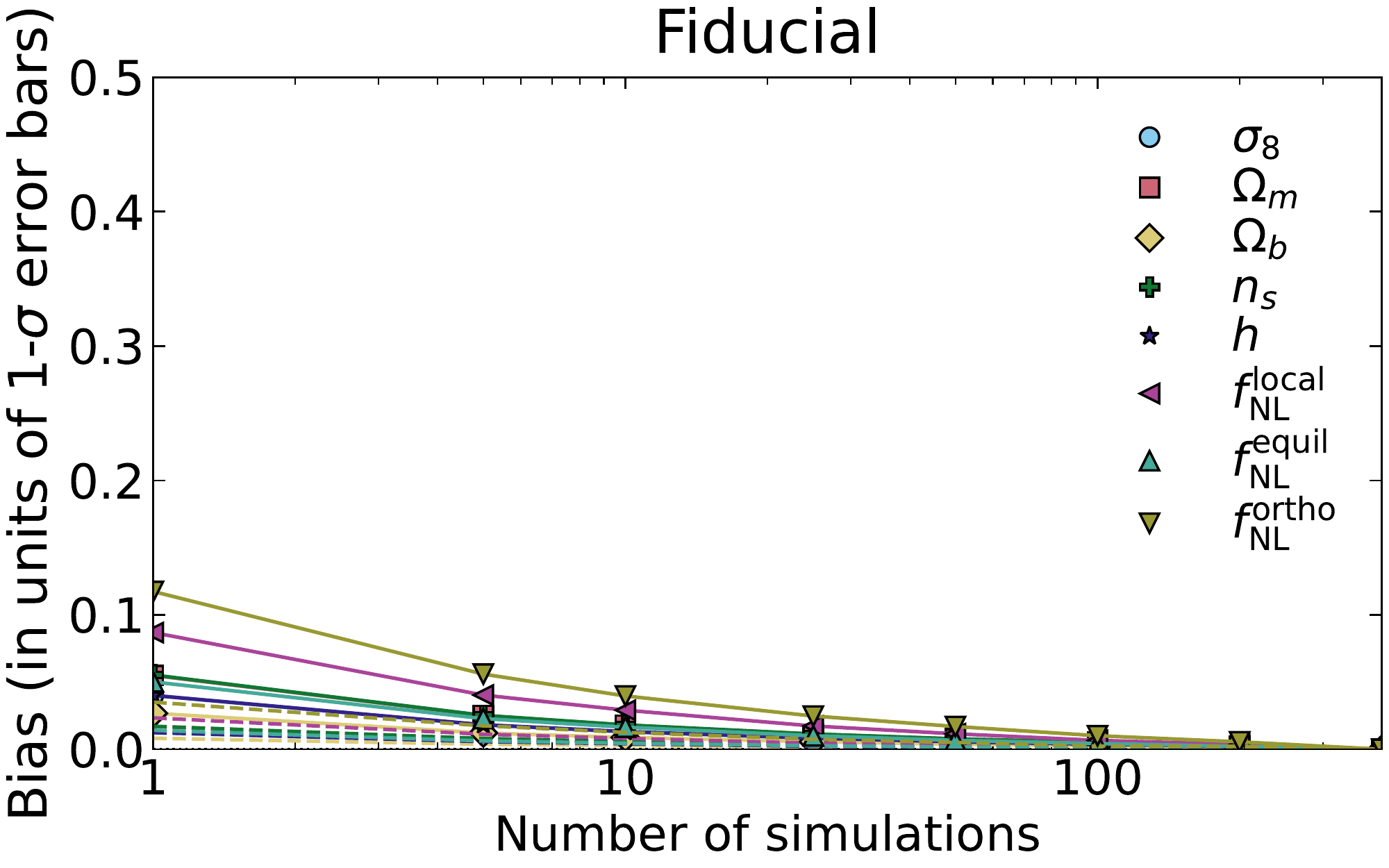}
    \includegraphics[width=0.33\linewidth]{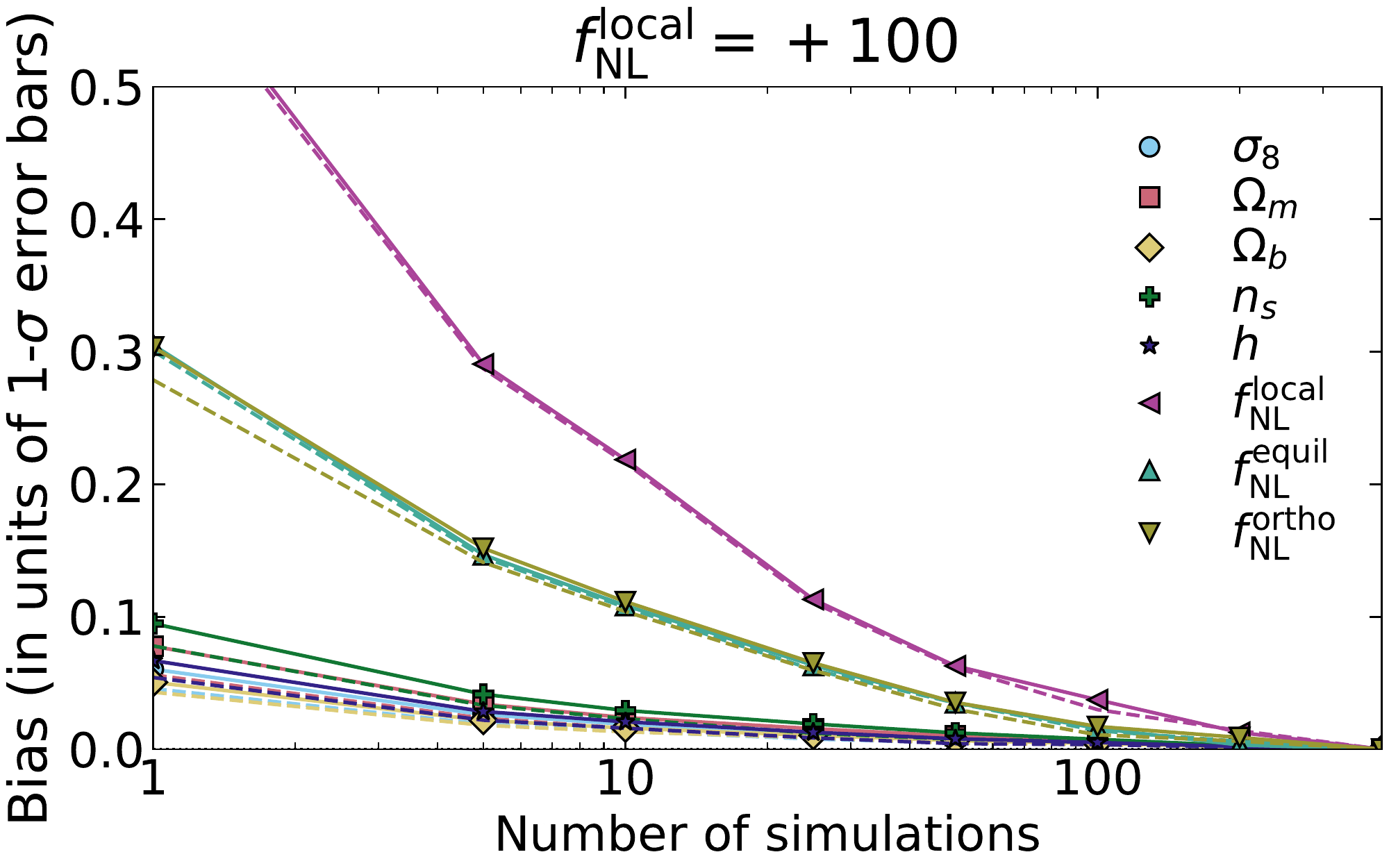}
    \includegraphics[width=0.33\linewidth]{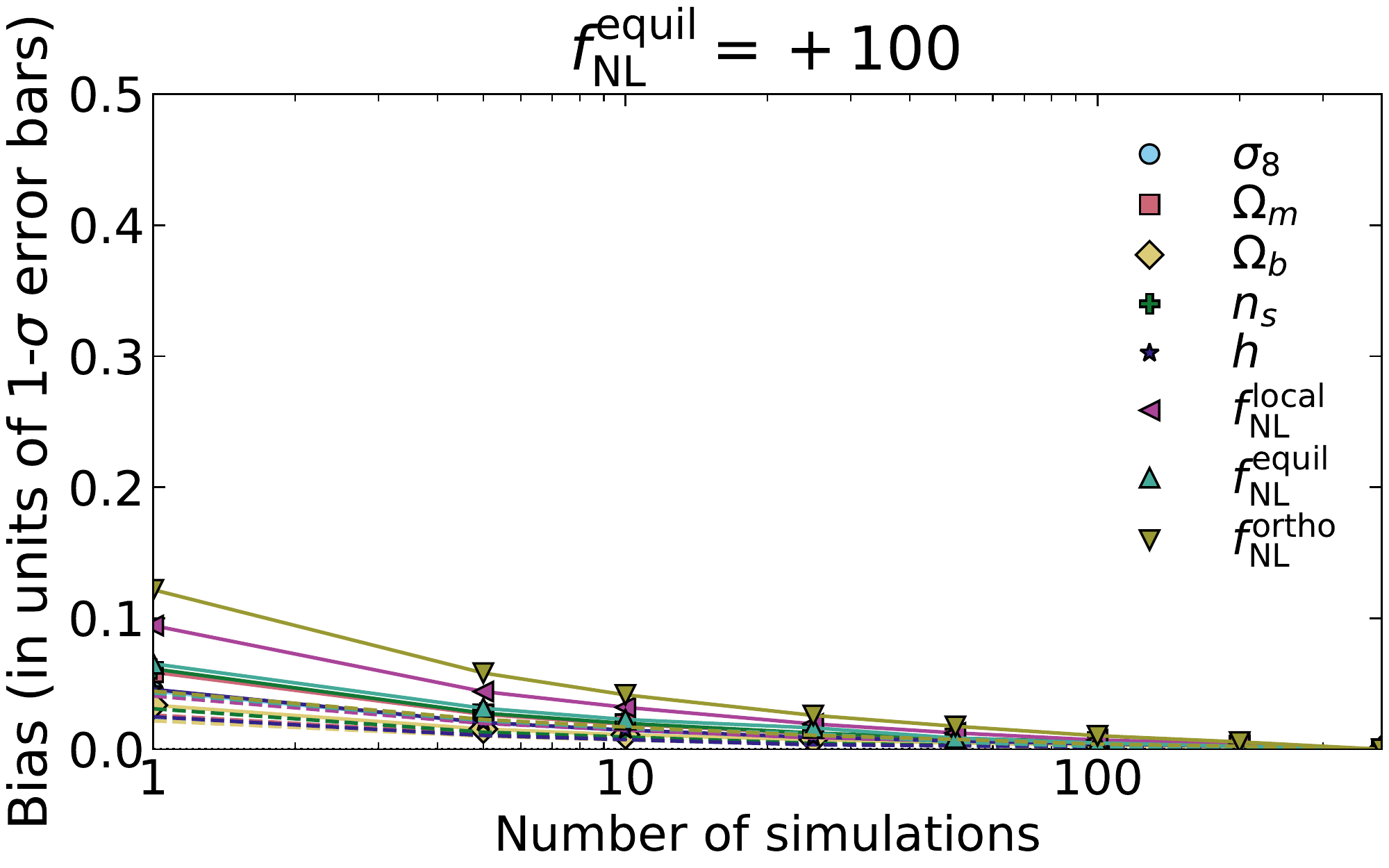}
    \caption{The typical size of the bias on estimated parameters, divided by their corresponding Fisher error bars, as a function of the number of simulations used to compute derivatives. We analyze \Quijote\ simulations for three different cosmologies, fiducial (left panel), $\fNLloc=100$ (middle panel) and $\fNLeq=100$ (right panel) and include both the power spectrum and the bispectrum up to $\kmax=0.5~\hMpc$. The solid lines correspond to datasets of 10 simulations, while the dashed lines correspond to datasets of 100 simulations.}
    \label{fig:derivatives-estimator}
\end{figure*}

For the derivatives, the impact on optimality is negligible, but the results can be biased, as shown in figure~\ref{fig:derivatives-estimator}. To study the typical size of the bias, we define less accurate estimators, using smaller subsets of the $500$ simulations, which are initially used to compute derivatives. For example, using $25$ simulations to compute derivatives gives 20 independent estimators. Then, we measure the difference between the measured parameters with every estimator and the benchmark results, obtained with an estimator calibrated using all available simulations. Finally, we compute the standard deviation of this difference as a function of the number of simulations used to calculate derivatives (from $1$ to $400$), which is what we show in figure~\ref{fig:derivatives-estimator} after dividing it by the corresponding Fisher error bar. 

We apply the same procedure for different sizes of data samples ($10$ and $100$ simulations) to verify that the results are indeed biased, and not suboptimal, as, in the latter case, the spurious bias evaluated with this methodology would decrease with the number of simulations. As data samples, we consider three different choices. In the first, all parameters are set at their fiducial values, meaning in particular that all PNG amplitudes are set to $0$. In the second, we set $\fNLeq=100$; we will define this case as ``close to fiducial'' since it corresponds to a $1$-$\sigma$ deviation from $\fNLeq=0$. In the third, we take $\fNLloc=100$; this is our ``far from fiducial'' case, since it is characterized by a $4$-$\sigma$ deviation from $\fNLloc=0$. We see that the impact of the number of simulations used to compute derivatives on the ``fiducial'' and ``close to fiducial'' cases is negligible. However, when analyzing the simulations with local PNG (``far from fiducial''), the measured value of $\fNLloc$ displays a bias. However, this effect decreases rapidly and becomes already negligible when using more than $100$ simulations to compute derivatives (let us recall that in our main analysis we use $500$ realizations for this task). Even when using less than $10$ simulations, it is still smaller than the typical $1$-$\sigma$ error bar. The two other primordial shapes, both being correlated with local PNG, are also affected in similar ways, whereas for cosmological parameters the effect is negligible. 

We can thus conclude that the quasi-maximum likelihood estimator combining power spectrum and modal bispectrum is unbiased and can optimally extract information about cosmology and PNG up to significantly non-linear scales, provided a sufficiently large number of simulations to calculate covariance and derivatives is used.

An important caveat is that, to achieve optimality, it is also important to evaluate covariances and derivatives by using an input fiducial cosmology in the simulations, which is close enough to the actual value of parameters in the data. In our current test, this is true by construction. However, in general, it may be necessary to reach a good enough fiducial set of parameters by iteration. This poses the numerical challenge of generating thousands of N-body simulations at several different cosmologies.  One efficient solution to overcome this issue is to use the CARPool method \citep{Chartier:2020pmu, Chartier:2021frd, Chartier:2022kjz}, which combines a few high-fidelity simulations with many fast surrogates (typically $100$-$1000$ times faster to produce) to obtain accurate estimates of the mean and covariance matrix of our observables. This approach is under investigation and will be further discussed in a forthcoming publication.
\section{Conclusions}
\label{sec:conclusions}

In this paper, we discussed the implementation of a new, simulation-based, joint power spectrum and bispectrum  statistical estimator of PNG amplitudes and standard $\Lambda$CDM parameters in LSS data. The methodology follows the general approach developed in \citet{Alsing:2017var, Heavens:1999am} and it is based on extracting summary power spectrum and bispectrum statistics from the data and computing the score function, making the reasonable assumption that sampling distributions are Gaussian to a good approximation. This produces a vector of optimally compressed statistics, which is then used to build a quasi-maximum likelihood estimator of the parameters. The required covariances and numerical derivatives with respect to parameters are evaluated from a large set of mock realizations with Gaussian and non-Gaussian initial conditions. To extract bispectrum summaries, we implement an efficient numerical pipeline for modal bispectrum estimation, allowing for a useful preliminary compression step, in which the information in all available Fourier triplets is described by a relatively small set of mode amplitudes. 

Our pipeline was applied to a large set of \Quijote\ simulations at $z=1$, including newly produced sets of realizations with PNG of the local, equilateral and orthogonal types. In the analysis presented here, we focused on the study of the dark matter field. Of course this is a vastly idealized scenario, compared to the analysis of the galaxy density field in redshift space, required in a real dataset. However, a preliminary study of the matter field is a very useful starting point, as it both enables to validate the entire methodology -- developed here for the first time -- and to address at the same time the interesting question of how much information on PNG and cosmological parameters we can in principle extract, by pushing the analysis to small scales, where perturbative approaches fail, as we show, even in this simplified context. To this purpose, we derived Fisher bounds for the parameter set $\{ \fNLloc, \fNLeq, \fNLort, \sigma_8, \Omega_m, \Omega_b, n_s, h\}$, using the covariance of the summary statistics -- extracted from a training set of $8000$ realizations -- at three different cut-off scales, namely $\kmax=0.07~\hMpc$, $\kmax=0.2~\hMpc$, $\kmax=0.5~\hMpc$. We then applied the quasi-maximum likelihood estimator to our mock data and verified its unbiasedness. 

The robustness of the results was thoroughly checked, for example by reducing the number of training realizations used to calibrate the estimator, or to compute the Fisher forecasts. We confirmed that all our results are fully converged and remain stable (e.g., with results within the $1$-$\sigma$ range, even when we used a few times less data to evaluate covariances). Moreover, we run the pipeline on a set of NG realizations, generated independently of the \Quijote\ suite and displaying a different input $\fNL$ from the training set; also in this case, we recovered the unbiasedness of the estimator. In further consistency tests, we considered power spectrum and bispectrum estimates separately. Interestingly, starting from bispectrum results, we found out that PNG constraints significantly improve when we add power spectrum estimates. This effects persists when we fix all standard cosmological parameters to their known fiducial values. This might seem surprising at first, since there are no $\fNL$ signatures at tree-level in the dark matter power spectrum, for any PNG shape. However, it turns out that the power spectrum works in this case as an ancillary statistic, able to improve the separation between the primordial (signal) and gravitational (``noise'') bispectrum component, through its correlation with the latter.

In summary, the outcome of our analysis shows that our pipeline can extract information on $\fNL$ and cosmological parameters up to small scales, deep into the non-linear regime, which is a very promising starting point. The overall estimation procedure is unbiased and fast, once input simulations to train the estimator have been generated. A large set of input realizations, based on the \Quijote\ suite, was produced to this purpose and will soon be made publicly available. These simulations are fully described in our companion paper \citep{Coulton:2022}. In the same companion paper, we also study in more detail the information content of the power spectrum and bispectrum, at $z=0$, addressing the important points of 
quantifying how large is the information contribution coming from non-linear scales and of establishing when such contribution saturates. This is done via a Fisher matrix analysis using bispectrum Fourier modes, which also includes a further battery of numerical convergence and validation tests.

In the work presented here, the training and validation sets are characterized in most cases by matching input cosmological parameters. In general, this would not be the case and too large a mismatch between the input fiducial cosmology and the actual value of parameters in the data would make the results suboptimal or biased. This can be solved by implementing a recursive procedure, in which the fiducial cosmology is updated by taking the best-fit values of the parameters at each step. Such an approach has the extra cost of having to generate new sets of simulations to recalibrate the estimator weights at each step. This issue can in principle be significantly alleviated by resorting to the CARPool method, which allows for accurate estimates of covariances by resorting to just a small set of high-fidelity realizations, combined with a large set of fast surrogates. This approach is under investigation and it will be the object of a separate publication.

In forthcoming works, we will also gradually extend our analysis to include more and more realistic scenarios: first we will analyze the actual density field of biased tracers (dark matter halos and galaxies) and later on we will account for redshift space, incomplete sky-coverage and other crucial observational effects, to finally reach the capability to analyze real datasets. Note that, given our simulation-based approach, this will not require any significant modification in the pipeline that was illustrated and validated here, but only changes in the input training sets and data.

\section*{Acknowledgements}

\noindent GJ, ML and MB were supported by the project "Combining Cosmic Microwave Background and Large Scale Structure data: an Integrated Approach for Addressing Fundamental Questions in Cosmology", funded by the MIUR Progetti di Ricerca di Rilevante Interesse Nazionale (PRIN) Bando 2017 - grant 2017YJYZAH. 

\noindent DK is supported by the South African Radio Astronomy Observatory (SARAO)
and the National Research Foundation (Grant No. 75415).

\noindent GJ, and ML also acknowledge support from the INDARK INFN Initiative (\url{https://web.infn.it/CSN4/IS/Linea5/InDark}), which provided access to CINECA supercomputing facilities (\url{https://www.cineca.it}).

\noindent MB acknowledges the use of computational resources from the parallel computing cluster of the Open Physics Hub (\url{https://site.unibo.it/openphysicshub/en}) at the Physics and Astronomy Department in Bologna.

\noindent LV acknowledges  ERC (BePreSySe, grant agree- ment 725327),  PGC2018-098866- B-I00 MCIN/AEI/10.13039/501100011033 y FEDER “Una manera de hacer Europa”, and the “Center of Excellence Maria de Maeztu 2020-2023” award to the ICCUB (CEX2019-000918-M funded by MCIN/AEI/10.13039/501100011033).

\noindent B.D.W. acknowledges support by the ANR BIG4 project, grant ANR-16-CE23-0002 of the French Agence Nationale de la Recherche; and the Labex ILP (reference ANR-10-LABX-63) part of the Idex SUPER, and received financial state aid managed by the Agence Nationale de la Recherche, as part of the programme Investissements d'avenir under the reference ANR-11-IDEX-0004-02.
The Flatiron Institute is supported by the Simons Foundation.

\appendix

\section{Modal estimator}
\label{app:modal}

\subsection{Modal basis}
\label{app:basis}

The custom modes we use in this work are based on those developed in \citet{Hung:2019ygc, Byun:2020rgl}. From the one-dimensional functions:
\begin{equation}
    \begin{split}
        &q_0^\mathrm{tree}=\sqrt{\frac{k}{P(k)}} \frac{5}{14},\\
        &q_1^\mathrm{tree}=\sqrt{\frac{k}{P(k)}} P_L(k),\\
        &q_2^\mathrm{tree}=-\sqrt{\frac{k}{P(k)}} P_L(k)k^2,\\
        &q_3^\mathrm{tree}=\sqrt{\frac{k}{P(k)}} \frac{P_L(k)}{k^2},\\
        &q_4^\mathrm{tree}=\sqrt{\frac{k}{P(k)}} \frac{3}{14}k^2,\\
        &q_5^\mathrm{tree}=\sqrt{\frac{k}{P(k)}} \frac{1}{14}k^4,\\
    \end{split}
    \label{eq:tree-basis}
\end{equation}
where $P_L(k)$ is the linear matter power spectrum, one can write the following four modes:
\begin{equation}
    \begin{split}
       &Q_0^\mathrm{tree}(k_1, k_2, k_3) = q_{\{0}^\mathrm{tree}(k_1) q_1^\mathrm{tree}(k_2) q_{1 \}}^\mathrm{tree}(k_3),\\
       &Q_1^\mathrm{tree}(k_1, k_2, k_3) = q_{\{0}^\mathrm{tree}(k_1) q_2^\mathrm{tree}(k_2) q_{3 \}}^\mathrm{tree}(k_3),\\
       &Q_2^\mathrm{tree}(k_1, k_2, k_3) = q_{\{1}^\mathrm{tree}(k_1) q_3^\mathrm{tree}(k_2) q_{4 \}}^\mathrm{tree}(k_3),\\
       &Q_3^\mathrm{tree}(k_1, k_2, k_3) = q_{\{3}^\mathrm{tree}(k_1) q_3^\mathrm{tree}(k_2) q_{5 \}}^\mathrm{tree}(k_3),\\
    \end{split}
    \label{eq:tree-modes}
\end{equation}
which fits exactly the standard tree-level matter bispectrum.

Similarly, the local bispectrum template given in eq.~\eqref{eqn:localbis} can be described by a single mode:
\begin{equation}
    Q_0^\mathrm{local}(k_1, k_2, k_3) = q_{\{0}^\mathrm{local}(k_1) q_1^\mathrm{local}(k_2) q_{1 \}}^\mathrm{local}(k_3),
    \label{eq:local-mode}
\end{equation}
where the one-dimensional basis functions are:
\begin{equation}
    \begin{split}
        &q_0^\mathrm{local}=\sqrt{\frac{k}{P(k)}} \sqrt{P_L(k)} k^{n_s/2 - 2},\\
        &q_1^\mathrm{local}=\sqrt{\frac{k}{P(k)}} \sqrt{P_L(k)} k^{2- n_s/2}.\\
    \end{split}
    \label{eq:local-basis}
\end{equation}

\subsection{Gamma matrix}
\label{app:gamma}

The gamma matrix is defined as the inner product (eq.\ \ref{eq:inner}) of the mode functions $Q_n$, given by:
\begin{equation}\label{eq:gamma_append}
    \gamma_{mn} \equiv\langle Q_m | Q_n \rangle 
    = \frac{1}{8\pi^4}\int_{\cal{V}_T} \d k_1 \d k_2 \d k_3 \, Q_m(k_1,k_2,k_3)Q_n(k_1,k_2,k_3).
\end{equation}
The above inner product matrix needs to be calculated only once for a chosen basis function $Q_n$ and scale range $[\kmin,\kmax]$. Different ways exist to calculate the integral of the $\gamma$ matrix, where these methods have been summarised and compared extensively in \citet{Byun:2020rgl}. In this work we use the method applied for the first time in \citet{KaragiannisPhd}, where the inner-product is calculated by utilising 1D FFTs. In order to describe this method, we need to take a step back and use the definition of the inner product used to derive eqs.\ \eqref{eq:beta} and \eqref{eq:M}, as well as eq.\ \eqref{eq:gamma_append}, which is: 
\begin{equation}\label{eq:inner_3d}
    \langle Q_m | Q_n \rangle =\int_{\bk_1}\int_{\bk_2}\int_{\bk_3} (2\pi)^3 \delta_D(\bk_{123})\frac{Q_m(k_1,k_2,k_3)Q_n(k_1,k_2,k_3)}{k_1k_2k_3},
\end{equation}
where for brevity we introduce the shorthand notations $\int_{\bk}\equiv\int \frac{\d^3k}{(2\pi)^3}$ and $\bk_{123}\equiv\bk_1+\bk_2+\bk_3$. Following the steps of \citet{Fergusson:2010ia}, we use
\begin{equation}
   \delta_D(\bk_{123})=\frac{1}{(2\pi)^3}\int \d^3x \, e^{i\bk_{123}\cdot \bx},
\end{equation}
where the exponent can be written as
\begin{equation}
   e^{i\bk\cdot \bx}=4\pi\sum_{\ell m} i^\ell j_\ell(k x)Y_{\ell m}(\hat{\bk})Y_{\ell m}^*(\hat{\bx}).
\end{equation}
Using the above equation into eq.\ \eqref{eq:inner_3d} we get

\begin{align}\label{eq:inner_3d_st2}
    \langle Q_m | Q_n \rangle= \int\d^3x \, (4\pi&)^3 \int_{\bk_1}\sum_{\ell_1 m_1} i^{\ell_1} j_{\ell_1}(k_1 x)Y_{\ell_1 m_1}(\hat{\bk}_1)Y_{\ell_1 m_1}^*(\hat{\bx}) \nonumber \\
    &\times\int_{\bk_2}\sum_{\ell_2 m_2} i^{\ell_2} j_{\ell_2}(k_2 x)Y_{\ell_2 m_2}(\hat{\bk}_2)Y_{\ell_2 m_2}^*(\hat{\bx}) \nonumber \\
    & \times\int_{\bk_3}\sum_{\ell_3 m_3} i^{\ell_3} j_{\ell_3}(k_3 x)Y_{\ell_3 m_3}(\hat{\bk}_3)Y_{\ell_3 m_3}^*(\hat{\bx}) \nonumber \\ & \times\frac{Q_m(k_1,k_2,k_3)Q_n(k_1,k_2,k_3)}{k_1k_2k_3}.
\end{align}
The integral over $\bk_i$ can be separated into a radial and angular part, where the latter is written as

\begin{equation}
    \int \d\Omega_{\bk_i} \, Y_{\ell_i m_i}(\hat{\bk}_i)=\sqrt{4\pi}\delta_{\ell_i0}\delta_{m_i 0},
\end{equation}
where to derive the above we have used the normalisation of spherical harmonics, i.e.\ $\int \d\Omega_{\bk_i} \, Y_{\ell_i m_i}(\hat{\bk}_i)^2=1$ and $Y_{00}=1/\sqrt{4\pi}$. The above equation forces all $\ell_i$ and $m_i$ in eq.\ \eqref{eq:inner_3d_st2} to zero, giving

\begin{align}\label{eq:inner_3d_st3}
    \langle Q_m | Q_n \rangle= \frac{(4\pi)^{9/2}}{(2\pi)^9}\int\d x\, x^2\int \d&\Omega_\bx  \int \d k_1 \, k_1^2j_0(k_1 x)Y_{00}(\hat{\bx}) \nonumber \\
    &\times\int \d k_2 \, k_2^2j_0(k_2 x)Y_{00}(\hat{\bx}) \nonumber \\
    & \times\int \d k_3 \, k_3^2j_0(k_3 x)Y_{00}(\hat{\bx}) \nonumber \\ & \times\frac{Q_m(k_1,k_2,k_3)Q_n(k_1,k_2,k_3)}{k_1k_2k_3}.
\end{align}
Now, if we use for the integration over $\bx$ the fact that, $\int \d \Omega_\bx \, Y_{00}(\bx)^3=1/\sqrt{4\pi}$ and $\int\d x \, x^2 j_0(k_1 x) j_0(k_2 x) j_0(k_3 x)=\pi/(8k_1k_2k_3)$, where the latter is non-zero only for wavevectors that form a triangle, then we retrieve the expression in eq.\ \eqref{eq:inner_3d}. However, at this point we will only use the expression for the angular part of the $\bx$ integral, while we substitute in eq.\ \eqref{eq:inner_3d_st3} the basis functions $Q_m=q_{\{p_1}(k_1) q_{r_1}(k_2) q_{s_1 \}}(k_3)$ and $Q_n=q_{\{p_2}(k_1) q_{r_2}(k_2) q_{s_2 \}}(k_3)$, which is a separable product of one-dimensional functions $q_p(k)$. Furthermore, we use the expression for the zero-order Bessel function, i.e.\ $j_0(k_ix)=\sin(k_ix)/(k_ix)$, thus getting

\begin{align}\label{eq:inner_3d_st4}
    \langle Q_m | Q_n \rangle = \frac{1}{2\pi^5}\int\d x \frac{1}{x} \frac{1}{6}\Big(&\xi_{p_1p_2}(x)[\xi_{r_1r_2}(x)\xi_{s_1s_2}(x)+\xi_{r_1s_2}(x)\xi_{s_1r_2}(x)] \nonumber \\
    & \xi_{p_1r_2}(x)[\xi_{r_1s_2}(x)\xi_{s_1p_2}(x)+\xi_{r_1p_2}(x)\xi_{s_1s_2}(x)] \nonumber \\
    & \xi_{p_1s_2}(x)[\xi_{r_1p_2}(x)\xi_{s_1r_2}(x)+\xi_{r_1r_2}(x)\xi_{s_1p_2}(x)]\Big).
\end{align}
where
\begin{equation}\label{eq:xi_1dFFT}
    \xi_{pr}(x)\equiv\int_{\kmin}^{\kmax} \d k\, q_{p}\left(\frac{k-\kmin}{\kmax-\kmin}\right)q_{r}\left(\frac{k-\kmin}{\kmax-\kmin}\right)\sin(kx).
\end{equation}
The above integral over $k$ can be calculated by using one-dimensional FFTs, as described in Chapter 13.9 of Numerical Recipes \citep{Numerical_Recipies}, while the integration over $x$ in eq.\ \eqref{eq:inner_3d_st4} can be done by using conventional numerical integration methods, like the five-point Newton–Cotes quadrature rule used here. The accuracy of the outlined method, on calculating eq.\ \eqref{eq:inner_3d_st4}, depends on the grid resolution of $k$, which has to be sufficient, not only to accurately calculate the integral of eq.\ \eqref{eq:xi_1dFFT}, but the derived range and resolution of $x$ generated by the FFT, should be adequate enough, as well, to retrieve accurate numerical result for the outer most integration over $x$. More details on this method can be found in \citep{Numerical_Recipies}.

\section{Impact of cosmological parameters}
\label{app:ratio-params}

\begin{figure}
    \centering
    \includegraphics[width=0.5\linewidth]{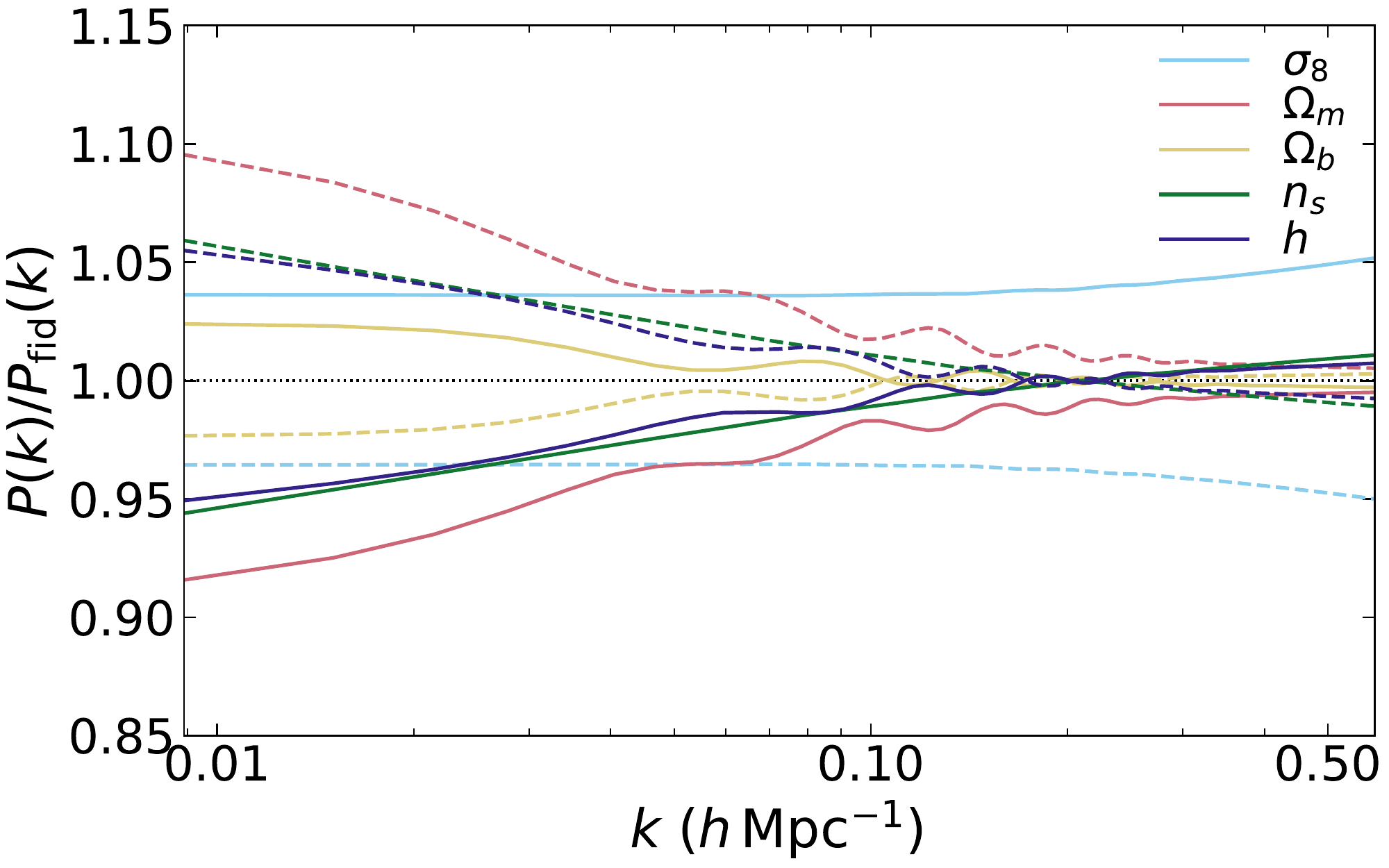}
    \caption{The impact of cosmological parameters on the matter power spectrum ( averaged from $500$ N-body simulations). Solid and dashed lines correspond respectively to values perturbed above and below the fiducial.}
    \label{fig:pk-cosmo}
\end{figure}

\begin{figure*}
    \includegraphics[width=\textwidth]{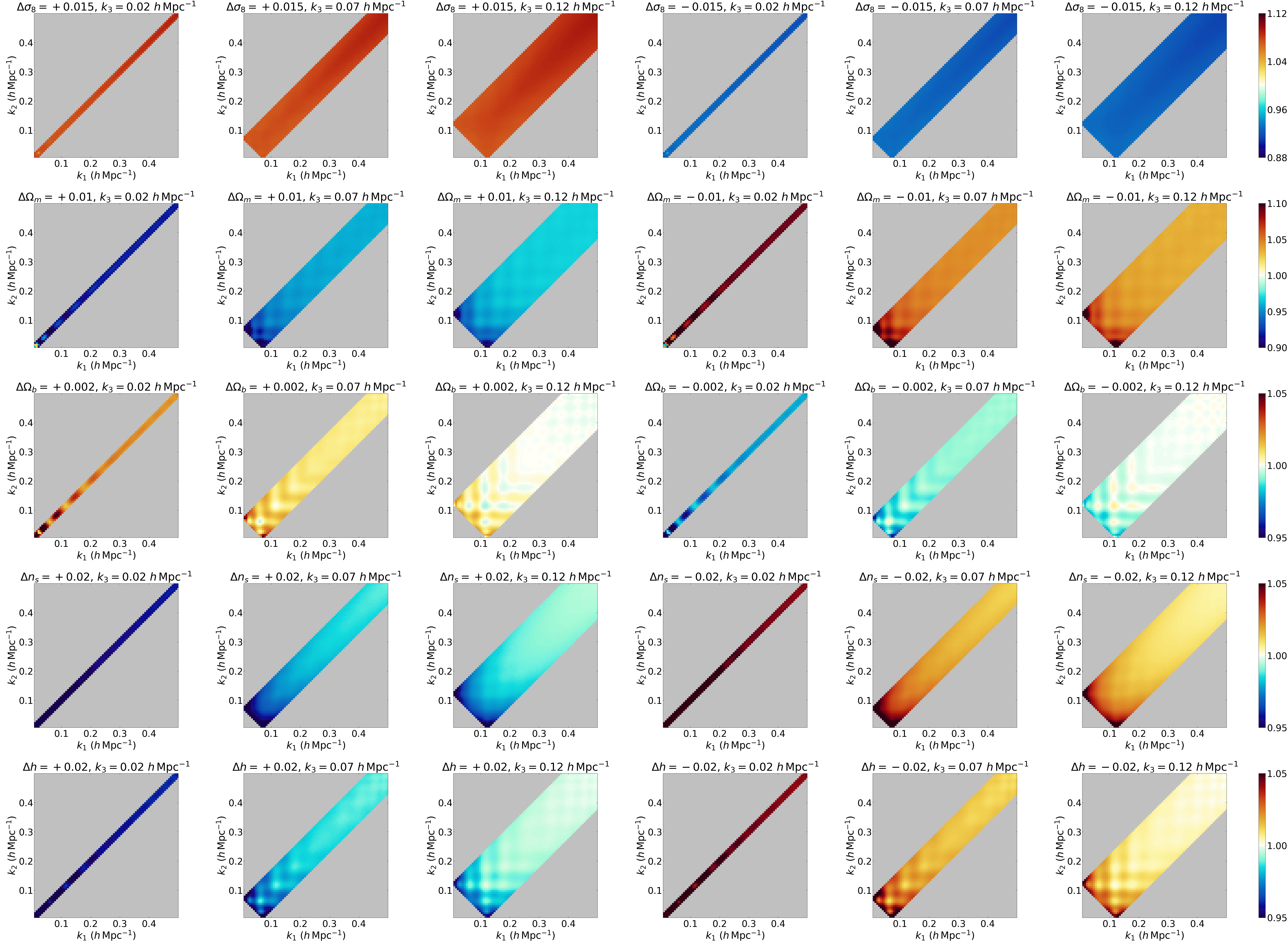}
    \caption{The impact of cosmological parameters on the matter bispectrum (ratio bispectrum with a modified parameter to fiducial, averaged from $500$ N-body simulations).}
    \label{fig:bisp-cosmo}
\end{figure*}

In figures~\ref{fig:pk-cosmo} and \ref{fig:bisp-cosmo}, we show the impact on the power spectrum and on the bispectrum of varying the cosmological parameters considered in this work, similar to what  was done for PNG in section~\ref{sec:measurements}. This highlights distinct behaviours for each parameter and each observable, in contrast with PNG which is strongly degenerate at the power spectrum level.

\bibliographystyle{aasjournal}
\bibliography{PNGmodalDM}

\end{document}